\documentclass[12pt]{article}
\usepackage{amsmath}
\usepackage{graphicx,psfrag,epsf}
\usepackage{enumerate}
\usepackage{natbib}
\usepackage{url} 

\newcommand{\blind}{0}

\usepackage[]{graphicx}
\usepackage[]{color}
\usepackage{alltt}
\usepackage{bm}
\usepackage{bbm}
\usepackage{amsmath, amsfonts}
\usepackage{natbib}
\usepackage{framed}
\usepackage{hyperref}

\usepackage{titlesec}
\titlespacing*{\section}{0pt}{10pt}{0.5\baselineskip}
\titlespacing*{\subsection}{0pt}{5pt}{0pt}
\titlespacing*{\subsubsection}{0pt}{5pt}{0pt}

\usepackage[doublespacing]{setspace}
\usepackage{tabularx} 
\usepackage{color, colortbl} 
\definecolor{Gray}{gray}{0.95}
\usepackage{booktabs} 
\newcolumntype{Y}{>{\centering\arraybackslash}X} 

\usepackage[font=small]{caption}

\usepackage{array}
\usepackage{geometry}
\newcolumntype{P}[1]{>{\centering\arraybackslash}p{#1}} 

\makeatletter
\def\maxwidth{ %
  \ifdim\Gin@nat@width>\linewidth
    \linewidth
  \else
    \Gin@nat@width
  \fi
}
\makeatother

\definecolor{fgcolor}{rgb}{0.345, 0.345, 0.345}

\makeatletter
 {\par\unskip\endMakeFramed%
 \at@end@of@kframe}
\makeatother

\definecolor{shadecolor}{rgb}{.97, .97, .97}
\definecolor{messagecolor}{rgb}{0, 0, 0}
\definecolor{warningcolor}{rgb}{1, 0, 1}
\definecolor{errorcolor}{rgb}{1, 0, 0}
\newenvironment{knitrout}{}{} 

\begin{document}

\date{}

\def\spacingset#1{\renewcommand{\baselinestretch}%
{#1}\small\normalsize} \spacingset{1}


\if0\blind
{
  \title{\bf Selective Inference for Additive and Linear Mixed Models}
  \author{
  David R\"ugamer\\ 
    Department of Statistics, LMU Munich\\
    and \\
    Philipp F.M. Baumann\\
    KOF Swiss Economic Institute, ETH Zurich\\
    and \\
    Sonja Greven\\
    School of Business and Economics, Humboldt-Universit\"at zu Berlin
    }
  \maketitle
} \fi

\if1\blind
{
  \bigskip
  \bigskip
  \bigskip
  \begin{center}
    {\LARGE\bf Title}
\end{center}
  \medskip
} \fi

\bigskip
\begin{abstract}
This work addresses the problem of conducting valid inference for additive and linear mixed models after model selection. One possible solution to overcome overconfident inference results after model selection is selective inference, which constitutes a post-selection inference framework, yielding valid inference statements by conditioning on the selection event. We extend recent work on selective inference to the class of additive and linear mixed models for any type of model selection mechanism that can be expressed as a function of the outcome variable (and potentially on covariates on which it conditions).
We investigate the properties of our proposal in simulation studies and  apply the framework to a data set in monetary economics. Due to the generality of our proposed approach, the presented approach also works for non-standard selection procedures, which we demonstrate in our application. Here, the final additive mixed model is selected using a hierarchical selection procedure, which is based on the conditional Akaike information criterion and involves varying data set sizes. 
\end{abstract}

\noindent%
\vfill

\newpage
\spacingset{1.45} 
\section{Introduction}

In practice, model selection is often done prior to hypothesis testing, or tests themselves are used for model selection. Very often, inference results are used without adjusting for preceding model selection, or iterative test procedures are based on uncorrected null distributions, thereby increasing the type I error notably \citep[see, e.g.,][]{Fithian.2014}. For mixed models, model selection is done in a variety of ways, e.g., by using the conditional Akaike information criterion \citep[cAIC,][]{Greven.2010,Saefken.2014,Saefken.2019} to select among different random effect structures or based on tests \citep[see, e.g.][]{lmerTest.2017}. Any of the used selection algorithms imply the necessity for adjusted inference post-model selection to avoid overconfident inference results. Automatic selection procedures that iteratively apply a statistical test for variable selection, in addition,  are based on an incorrect distribution assumption in all but the first step and thus may yield undesirable results.

To overcome overconfident inference results after model selection, a variety of post-selection inference approaches exist. These adjust classical inference for the additional stochastic aspect in selecting the model and hence the tested hypothesis. Sparked by various proposals such as \citet{Berk.2013}, many authors have addressed the problem of valid inference after model selection in different ways. \citet{Berk.2013} establish an inference concept, which is independent of the used model selection procedure and often referred to as \emph{Post-Selection Inference} (or short \emph{PoSI}). Another research direction in valid post-selection inference focuses on \textit{selective inference}, a post-model selection procedure that motivates a range of methods that target specific selection algorithms, especially the Lasso \citep[see, e.g.,][]{Fithian.2014, Lee.2016, Tibshirani.2016}. We focus here on the concept of selective inference. Selective inference allows for conducting valid inference conditional on a certain selection event by adjusting the distribution of widely used test statistics. Most of the proposed selective inference methods, however, only deal with the linear model.

\paragraph{Our contribution.} In this work we close an important gap in the application of linear and additive mixed models (LMMs). In particular, we extend the selective inference framework to the class of linear and additive mixed models and show how to allow for any type of model selection mechanism re-applicable on new data.
This allows us to develop valid inference post model selection  for  fixed and random effects as well as linear combinations of these effects in LMMs.
Our work thereby contributes to and extends research on selective inference by a) incorporating random effects estimated with shrinkage, b) extending the ordinary least squares-type test statistics used in selective inference and providing results for post-selection inference based on marginal and conditional distributions in mixed models, c) proposing and evaluating different approaches to overcome the assumption of known error and random effect covariance matrices, d) transferring our approach to selective inference for additive (mixed) models and e) adopting recent ideas on Monte Carlo approximation to allow for inference in these models after any model selection criterion that can be stated as a deterministic function of the response. We additionally make available an R package \texttt{selfmade} (SELective inFerence for Mixed and Additive model Estimators) 
and all codes for simulations on Github  (\url{https://github.com/davidruegamer/selfmade}).

In the following, we summarize the theory of selective inference for linear models in Section~\ref{sec.SI} and extend existing approaches for selective inference to mixed models in Section~\ref{sec.MM}. 
We further derive a selective inference concept for additive models in Section~\ref{sec.AM} and present simulation results for the proposed methods in Section~\ref{sec.sim}. We apply our results to the world inflation data, compare results with existing approaches in Section~\ref{sec.appl} and summarize our concept in Section~\ref{sec.disc}.

\section{Selective Inference} \label{sec.SI}

We start with briefly summarizing selective inference foundations in \citet{Loftus.2015a,Yang.2016,Ruegamer.2018c} and describe different concepts for the linear model. We will therefore introduce our notation which is later extended for linear mixed models. Let $\bm{Y} = (Y_1, \ldots, Y_n)^\top$ be a vector of $n$ independent random variables following $\bm{Y} | \bm{X} \sim \mathcal{N}(\bm{\mu}, \sigma^2 \bm{I}_n)$ with mean vector $\bm{\mu} = (\mu_1, \ldots, \mu_n)^\top$, variance $\sigma^2$, $\bm{I}_n$ the $n$-dimensional identity matrix and $\bm{X}$ a $n\times p$ matrix of $p$ covariate columns $\bm{x}_1, \ldots, \bm{x}_p \in \mathbb{R}^n$. We are interested in estimating $\bm{\mu}$, the conditional expectation of $\bm{Y}$, which we model by a linear combination $\bm{X}_{\mathcal{A}} \bm{\beta}_{\mathcal{A}}$ of a subset $\bm{X}_{\mathcal{A}}$ of the data repository $\bm{X}$, where $\mathcal{A} \subset \mathcal{P}(\{1,\ldots,p\})$ is an column indexing set and $\mathcal{P}(A)$ denotes the power set of a set $A$. One key aspect of selective inference is that no assumption on the true underlying mean structure is made. Instead the true mean $\bm{\mu}$ can have an arbitrary structure, can be potentially non-linear in observed covariates $\bm{X}$ and may depend on unobserved covariates not contained in $\bm{X}$. The inference goal after selection of some working structure $\bm{X}_{\mathcal{A}} \bm{\beta}_{\mathcal{A}}$ from the set of all potential effects based on $\bm{X}$ is to infer about $\bm{\beta}_{\mathcal{A}}$, the coefficients for the projection of $\bm{\mu}$ onto the column space of $\bm{X}_{\mathcal{A}}$. However, we would like to correct resulting inference statements for the fact that we have selected the set $\mathcal{A}$ of covariates with some selection procedure $\mathcal{S}$ depending on the data $\bm{Y}$, i.e.\ $\mathcal{A}:=\mathcal{S}(\bm{y})$ for the observed realization $\bm{y}$ of $\bm{Y}$. Note that this selection procedure usually implicitly depends on and conditions on the given data $\bm{X}$, but we will suppress this in the notation in the following for better readability. In particular, we are usually interested in the effect of one specific direction $j$ of the projection $\bm{X}_{\mathcal{A}} \bm{\beta}_{\mathcal{A}}$, which is denoted by $\beta_{\mathcal{A}_j} = \bm{e}_j^\top \bm{\beta}_{\mathcal{A}}$ with $j$th unit vector $\bm{e}_j$. We call this the $j$th \emph{selective direction effect} or short \emph{SDE} in the following. 
SDEs and corresponding quantities are now discussed in more detail in the light of defined hypotheses and test statistics.

\subsection{Test Statistic and Test Vector}

We now define a general hypothesis and corresponding test statistic for either testing a single coefficient or testing a group of coefficients. When testing a group of coefficients we are interested in a projection $\bm{P}_{\bm{W}}$ of  $\bm{\mu}$ onto a linear $\text{span}(\bm{W}) \subset \mathbb{R}^n$ with $\bm{W} \in \mathbb{R}^{n\times w}, w\in \mathbb{N}$, $w > 1$, and our null hypothesis can be stated as
\begin{equation}
H_0: || \bm{P}_{\bm{W}} \bm{\mu}||_2 = \tilde{\rho}
\end{equation}
where $\tilde{\rho}$ is the value assumed under the null hypothesis and $||\cdot||_2$ is the Euclidean norm \citep{Yang.2016,Ruegamer.2018c}. We define the test statistic as 
\begin{equation}
\tilde{T} = ||\bm{P}_{\bm{W}} \bm{Y} ||_2. \label{eq:teststat}
\end{equation}
We follow \citet{Loftus.2015a, Yang.2016, Ruegamer.2018c} and define $\bm{W} = \bm{P}^\bot_{\bm{X}_{\mathcal{A}\backslash j}} \bm{X}_j$ with $\bm{X}_{\mathcal{A}\backslash j}$ denoting the selected design matrix without the column(s) $\bm{X}_j$ corresponding to the $j$th selected (group) variable and $\bm{P}^\bot_{\bm{X}_{\mathcal{A}\backslash j}}$ corresponding to the projection onto the orthogonal complement of the subspace spanned by ${\bm{X}_{\mathcal{A}\backslash j}}$. This corresponds to testing whether the $j$th group variable in $\mathcal{A}$ is correlated with $\bm{\mu}$ after correcting for all other covariates in the model $\mathcal{A}$.

When testing single coefficients, $w = 1$, $\bm{W} = \bm{v} \in \mathbb{R}^{n \times 1}$ is a vector and $\bm{P}_{\bm{W}} = \bm{P}_{\bm{v}}$. Alternatively, a signed and unscaled null hypotheses 
\begin{equation}
H_0: \bm{v}^\top \bm{\mu} = \rho
\end{equation} 
and test statistic
\begin{equation} \label{eq:unscaled}
T = \bm{v}^\top \bm{Y}
\end{equation}
can be defined, where $\bm{v}^\top \bm{\mu}$ corresponds to the $j$th selective direction effect in linear models $\bm{e}_j^\top \bm{\beta}_{\mathcal{A}} = \bm{e}_j^\top (\bm{X}_{\mathcal{A}}^\top \bm{X}_{\mathcal{A}})^{-1} \bm{X}_{\mathcal{A}}^\top \bm{\mu}$, namely the $j$th component of the coefficient vector $\bm{\beta}_{\mathcal{A}}$ in the projection $\bm{X}_{\mathcal{A}} \bm{\beta}_{\mathcal{A}}$ of $\bm{\mu}$ into $\text{span}(\bm{X}_{\mathcal{A}}) \subset \mathbb{R}^n$, and $\bm{v}^\top := \bm{e}_j^\top (\bm{X}_{\mathcal{A}}^\top \bm{X}_{\mathcal{A}})^{-1} \bm{X}_{\mathcal{A}}^\top$, which yields $\hat{\beta}_{\mathcal{A},j} = \bm{v}^\top \bm{Y} = T$. 

\subsection{Inference Region}

Given the direction(s) of interest $\bm{v}$ or $\bm{W}$, we are interested in the distribution of $T$ or $\tilde{T}$ conditional on the selection event $\mathcal{S}(\bm{Y}) = \mathcal{A}$. For many selection procedures, the test statistic is restricted by the selection event to some set $\mathcal{T}$, i.e. $T$ has to lie in $\mathcal{T}$ conditional on $\mathcal{S}(\bm{Y}) = \mathcal{A}$. If $T$ follows a Gaussian distribution and model selection is, e.g., performed by the Lasso, the restricted space can be described explicitly \citep[see, e.g.,][]{Lee.2016, Tibshirani.2016}, $Y$ turns out to lie in a polyhedron and $\mathcal{T}$ is an interval in this polyhedron. $T$ thus follows a truncated normal distribution with support $\mathcal{T}$ when conditioning on the selection event. In other cases, the selection induces an inference region $\mathcal{T}$ that cannot be characterized mathematically \citep{Yang.2016}, or while it
can be characterized, calculating the resulting truncation for $T$ is not feasible \citep{Ruegamer.2018c}. A possible solution for these cases is to numerically explore the region $\mathcal{T}$ restricting the test statistic and its distribution using Monte Carlo sampling \citep{Yang.2016, Ruegamer.2018c}.

\subsection{Monte Carlo Approximation}\label{MC}

Monte Carlo approximation for selective inference has been proposed by several authors \citep[see, e.g.,][]{Fithian.2014, Yang.2016, Ruegamer.2018c}. For the linear model with known variance $\sigma^2$, observe that we can decompose $\bm{Y}$ as $\tilde{T} \cdot \text{dir}_{\bm{W}}(\bm{Y}) + \bm{\mathfrak{Z}}$ with $\text{dir}_{\bm{W}}(\bm{Y}) = \frac{\bm{P}_{\bm{W}} \bm{Y}}{||\bm{P}_{\bm{W}} \bm{Y} ||_2}$ and $\bm{\mathfrak{Z}} = \bm{P}^\bot_{\bm{W}} \bm{Y}$. Alternatively, for the unscaled version (\ref{eq:unscaled}), we can decompose $\bm{Y}$ as $\frac{\bm{v}}{\bm{v}^\top \bm{v}} \cdot T + \bm{\mathfrak{Z}}$ and $\bm{\mathfrak{Z}} = \bm{P}_{\bm{v}}^\bot \bm{Y}$ with $\bm{P}_{\bm{v}}^\bot = \bm{I}_n - \frac{\bm{v}\bm{v}^\top}{\bm{v}^\top \bm{v}}$. When conditioning on the selection event $\mathcal{S}(\bm{Y}) = \mathcal{A}$, where $\mathcal{A}:= \mathcal{S}(\bm{y})$ for the given realization $\bm{y}$, and when additionally conditioning on $\text{dir}_{\bm{W}}(\bm{Y}) =\text{dir}_{\bm{W}}(\bm{y})$, the contribution of the SDE can be quantified by the magnitude of $\tilde{T}$ \citep{Yang.2016}. When we additionally condition on $\bm{\mathfrak{Z}} = \bm{P}^\bot_{\bm{W}} \bm{y} =: \bm{\zeta}$, the only variation of $\bm{Y}$ left is in $\tilde{T}$. An important special case is given for $w \geq 2$ and $\tilde{\rho} = 0$, i.e., when testing the significance of the SDE of a group. When conditioning on $\mathcal{S}(\bm{Y}) = \mathcal{A}, \text{dir}_{\bm{W}}(\bm{Y}) =\text{dir}_{\bm{W}}(\bm{y})$ and $\bm{\mathfrak{Z}} = \bm{\zeta}$, we have $\tilde{T} \sim \sigma \cdot \chi_w$ restricted to $\mathcal{T}$. If $w \geq 2$ and $\tilde{\rho} \neq 0$, \citet{Yang.2016} allow for computation of p-values by deriving $f_{\tilde{T}}$, the (conditional) density of $\tilde{T}$ (conditional on the selection event, $\bm{\mathfrak{Z}} = \bm{\zeta}$ and $\text{dir}_{\bm{W}}(\bm{y})$), and rewriting the p-value $P$ for the observed value $\tilde{T} = \tilde{t}_{\text{obs}}$ as a ratio of two expectations, which can be approximated numerically.

Along the same lines \citet{Ruegamer.2018c} use this idea to calculate p-values and confidence intervals (CIs) for a normally distributed test statistic $T$ with potentially multiple truncation limits. In this case 
$T \sim \mathcal{N}(\rho,\sigma^2 \bm{v}^\top \bm{v} =: \kappa)$ without restriction and after selection, $T$ follows a truncated version of this normal distribution with the same moments but support $\mathcal{T}$. P-values and CIs can be calculated using 
\begin{equation}
\varsigma(\rho) = \frac{\int_{\mathcal{T}, t > t_{\text{obs}}} \exp \{ -\kappa^{-1} (t-\rho)^2 / 2\} \,dt}{\int_{\mathcal{T}} \exp \{ -\kappa^{-1} (t-\rho)^2 / 2\} \,dt} = \frac{\mathbb{E}_{T \sim \mathcal{N}(0,\kappa)}[ e^{T\cdot \rho / \kappa} \mathbbm{1}\{ T \in \mathcal{T}, T > t_{\text{obs}} \} ]}{\mathbb{E}_{T \sim \mathcal{N}(0,\kappa)}[ e^{ T\cdot \rho / \kappa} \mathbbm{1} \{ T \in \mathcal{T}\} ]}, \label{eq:pvalYang}
\end{equation}
corresponding to $H_0: \bm{v}^\top \bm{\mu} = \rho$ for testing a certain null hypothesis or, by inverting the test, for CIs by searching for lower and upper interval bounds $(\rho_{\alpha/2},\rho_{1-\alpha/2})$, $\alpha \in (0,1)$ such that $\varsigma(\rho_a) = a, a \in \{\alpha/2,1-\alpha/2\}.$

By drawing $B$ samples $T^b$, $b = 1,\ldots, B$, of $T$ from the normal distribution (or samples $\tilde{T}^b$ of $\tilde{T}$ from a $\chi$-distribution), we can check the congruency with the initial selection given by $\mathcal{S}(\bm{y}) = \mathcal{A}$ by defining $\bm{Y}^b := \frac{\bm{v}}{\bm{v}^\top \bm{v}} T^b + \bm{\zeta}$ (or $\bm{Y}^b := \tilde{T}^b \cdot \text{dir}_{\bm{W}}(\bm{y}) + \bm{\zeta}$ for the grouped effect test) and evaluating $\mathcal{S}(\bm{Y}^b)$. Given enough samples $T^b \in \mathcal{T}$, p-values can be approximated by replacing the expectations in (\ref{eq:pvalYang}) by sample averages. Since the survival function of $T$ is monotone in its mean, we can furthermore invert the hypothesis test to construct a selective confidence interval as proposed in \citet{Yang.2016, Ruegamer.2018c}. In cases where the observed test statistic lies in an area where the null distribution has little probability mass, empirical approximation of (\ref{eq:pvalYang}) might be difficult or sampling might even yield only values outside the inference region of interest for realistic values of $B$. We therefore adapt and extend the approaches by \citet{Yang.2016, Ruegamer.2018c} and use an importance sampler with proposal distribution $\mathcal{N}(t_{\text{obs}}, \kappa)$ or a mixture of normal distributions with different locations.

\section{Selective Inference for Linear Mixed Models} \label{sec.MM}

For the analysis of longitudinal or clustered data, the linear mixed model is a natural choice. In addition to a wide range of linear mixed models (LMMs), any additive model incorporating a quadratic penalty can be framed as a mixed model \citep{Wood.2017}, further extending the LMM model class. In the following, we extend existing approaches for selective inference in linear models to the class of linear mixed models. We use the following linear mixed model notation:
\begin{equation}
\bm{Y} = \bm{X}\bm{\beta} + \bm{Z}\bm{b} + \bm{\varepsilon}, \label{LMM}
\end{equation}
with residual term $\bm{\varepsilon} \sim \mathcal{N}(\bm{0}, \bm{R})$ and random effects $\bm{b} \sim \mathcal{N}(\bm{0}, \bm{G})$, $\bm{\varepsilon} \bot \bm{b}$ with corresponding covariance matrices $\bm{R}$ and $\bm{G}$, respectively. In the following, this setup is referred to as the \emph{working model} as we do not necessarily assume the true data generating process to be of the form (\ref{LMM}). 

When adapting the principles of selective inference for LMMs, some of the prerequisites for the framework proposed by \citet{Lee.2016, Yang.2016, Ruegamer.2018b} remain the same, whereas several key aspects change and thus extensions of the previous works are required. First and foremost, LMMs come with different distributional viewpoints. Inference and model selection can be conducted based either on the marginal distribution 
$\bm{Y} \sim \mathcal{N}(\bm{X}\bm{\beta}, \bm{\Sigma}) \label{LMMmarg}$
with $\bm{\Sigma} = \bm{Z}\bm{G}\bm{Z}^\top + \bm{R}$ or based on the conditional distribution $\bm{Y}|\bm{b} \sim \mathcal{N}(\bm{X}\bm{\beta} + \bm{Z}\bm{b}, \bm{R})$.
We here present two different ways to conduct selective inference for LMMs based on the marginal and on the conditional perspective. 
In contrast to most of the literature on selective inference, both approaches are not restricted to specific model selection criteria but only require the model selection procedure to be 
a function of the outcome variable (but can also depend on the covariates). In other words, our method provides selective inference for any selection procedure $\mathcal{S}: \bm{y} \to \mathcal{A}$ that is deterministic in $\bm{y}$. Note that, as for the linear model, the selection procedure implicitly conditions on the covariates (here $\bm{X}$ and $\bm{Z}$), which we will omit in the notation for better readability.

\paragraph{Practical Relevance.} Before introducing the methods in more detail, we want to first highlight three important practical aspects of the presented methods. 1) The following approaches assume to some extent knowledge of involved covariance matrices. We explicitly address this limitation in Section~\ref{sec.limits} and show in our numerical studies that the proposed solution yields valid inference in practice. 2) The purpose of the marginal perspective in Section~\ref{marg} lies in the selection of fixed effects and assumes that the covariance is known. 3) The conditional perspective in Section~\ref{cond} can also be used if random effect selection is of interest. This perspective is accompanied with less strict assumptions and is usually the recommended approach. In general, the choice of either one of these perspectives is strongly related to the assumptions on the underlying data generating process \citep[see, e.g.,][for a discussion on the use of the marginal and conditional perspective in linear mixed and additive models]{Greven.2010}.

\subsection{Marginal Perspective} \label{marg}

We first discuss the marginal perspective, which exhibits strong connections to selective inference concepts proposed for linear models (LMs).
When the question of interest -- and the focus of model selection -- are the fixed effects $\bm{\beta}$ of the LMM, the marginal distribution of $\bm{Y}$ is typically used to conduct inference. 
We can then utilize the framework of \citet{Lee.2016} and others, which provides selective inference statements in LMs for a normally distributed response $\bm{Y}$ with potentially non-diagonal variance-covariance matrix $\bm{\Sigma}$.

\subsubsection{Setup, Assumptions and Inference Goal} \label{margsetup}

We first assume a known covariance structures $\bm{\Sigma}$ given by a corresponding mixed model with known random effects covariance $\bm{G}$, random effects structure $\bm{Z}$ and error covariance $\bm{R}$. We will discuss unknown covariance structures in Section~\ref{sec.limits}. Overall, assume that \begin{equation}\label{marginalLMM}
\bm{Y} \sim \mathcal{N}(\bm{\mu}, \bm{ZG}\bm{Z}^\top + \bm{R} =: \bm{\Sigma})
\end{equation}
which is implied by $\bm{Y}|\bm{b} \sim \mathcal{N}(\bm{\mu} + \bm{Z}\bm{b}, \bm{R})$, as well as the working model as defined in (\ref{LMM}). $\bm{\mu}$ is allowed to have a flexible structure, potentially incorporating effects of unobserved covariates or non-linear effects of $\bm{X}$. After selection of a linear fixed effects (working) structure $\bm{X}_{\mathcal{A}} \bm{\beta}_{\mathcal{A}}$ from the set of all potential fixed effects based on $\bm{X}$, our goal is to infer about the $j$th selective direction effect given a fixed covariance structure $\bm{\Sigma}$:
\begin{equation}
H_0: \beta_{\mathcal{A}_j} = \beta_{j0} \label{fixedH0}.
\end{equation}
Alternatively, we may want to infer about a group of coefficients as introduced in Section~\ref{sec.SI}.

\subsubsection{Null Distribution and Test Statistic} \label{margteststat}

In line with the previous works on selective inference in LMs, we first assume that $\bm{\Sigma}$ is known. We are still interested in testing $H_0: \bm{v}^\top \bm{\mu} = \rho$, but in addition to the test statistic $T = \bm{v}^\top \bm{Y}$ with $\bm{v} = \bm{X}_{\mathcal{A}} (\bm{X}_{\mathcal{A}}^\top \bm{X}_{\mathcal{A}})^{-1} \bm{e}_j$ used in LMs \citep[see, e.g.,][]{Ruegamer.2018b} with null distribution $\mathcal{N}(\rho,\bm{e}_j^\top (\bm{X}_{\mathcal{A}}^\top \bm{X}_{\mathcal{A}})^{-1}
(\bm{X}_{\mathcal{A}}^\top \bm{\Sigma} \bm{X}_{\mathcal{A}}) (\bm{X}_{\mathcal{A}}^\top \bm{X}_{\mathcal{A}})^{-1} \bm{e}_j)$ in the setting of \eqref{marginalLMM}, we propose another test statistic $T_{\text{eff}} := \bm{v}_{\text{eff}}^\top \bm{Y}$ with $\bm{v}_{\text{eff}} = \bm{\Sigma}^{-1} \bm{X}_{\mathcal{A}} (\bm{X}_{\mathcal{A}}^\top \bm{\Sigma}^{-1} \bm{X}_{\mathcal{A}})^{-1} \bm{e}_j$ and null distribution $\mathcal{N}(\rho,\bm{e}_j^\top(\bm{X}_{\mathcal{A}}^\top \bm{\Sigma}^{-1} \bm{X}_{\mathcal{A}})^{-1} \bm{e}_j)$. $T_{\text{eff}}$ corresponds to the more efficient estimator $\hat{\bm{\beta}}_{\text{eff}} = \bm{v}_{\text{eff}}^\top \bm{Y}$ in the presence of a non-diagonal variance-covariance matrix for $\bm{Y}$. Like $T$, the test statistic has the desired property $\mathbb{E}(T_{\text{eff}}) = \beta_j$ in the case, in which the true mean $\bm{\mu}$ is a linear combination of $\bm{X}_{\mathcal{A}}$. Similarly to before, using this test statistic we can also decompose $\bm{Y}$ into $\bm{P}_{v_\text{eff}} \bm{Y}$ in the direction of $\bm{v}_{\text{eff}}$ with $\bm{P}_{\bm{v}_\text{eff}} = \bm{\Sigma} \bm{v}_\text{eff}\bm{v}_\text{eff}^\top / (\bm{v}_\text{eff}^\top \bm{\Sigma} \bm{v}_\text{eff})$ and an orthogonal complement $\bm{P}_{\bm{v}_\text{eff}}^\bot \bm{Y}$ such that $\text{Cov}(\bm{P}_{\bm{v}_\text{eff}} \bm{Y}, \bm{P}_{\bm{v}_\text{eff}}^\bot \bm{Y}) = \bm{0}$. This allows us to create new test statistics $T_\text{eff}^b$ and corresponding response values $\bm{Y}^b = \bm{\Sigma} \bm{v}_\text{eff} / (\bm{v}_\text{eff}^\top \bm{\Sigma} \bm{v}_\text{eff}) \cdot T^b_\text{eff} + \bm{P}_{\bm{v}_\text{eff}}^\bot \bm{y}$ for the observed $\bm{y}$ that only vary in the direction of interest defined by $\bm{v}_{\text{eff}}$. In contrast to the test vector $\bm{v}$ proposed for LMs, $\bm{P}_{\bm{v}_\text{eff}}$ in this case depends on $\bm{G}$.



\subsection{Conditional Perspective} \label{cond}

We now turn to the conditional linear mixed model perspective, where some assumptions can be relaxed, and which additionally allows for inference for additive models as presented in Section~\ref{sec.AM}.

\subsubsection{Setup, Assumptions and Inference Goal} \label{condsetup}

Assume that $\bm{Y} \sim \mathcal{N}(\bm{\psi}, \bm{R})$ and 
first relax the assumption of having a fixed random effect structure $\bm{Z}$ as well as known random effects covariance structure $\bm{G}$ that was used for the marginal perspective. We will use a working mixed model to model the (conditional) expectation $\bm{\psi}$, where $\bm{\psi}$ may or may not incorporate random effects. If $\bm{\psi}$ is of the form as assumed in \eqref{LMM} in Section \ref{margsetup}, we can also derive a corresponding marginal model. In the case where $\bm{\psi}$ incorporates random effects, the conditional approach is more suitable if the random effect structure is not known beforehand and, in particular, if the random effects are part of the model selection. We denote the set of selected random effects out of all potential random effect candidates by $\bm{b}_{\mathcal{B}}$ with $\mathcal{B}$ the corresponding index set.
Let the goal of the analysis be to infer about the selective direction effect $\bm{\beta}_{\mathcal{A}} \in \mathbb{R}^p$ or $\bm{b}_{\mathcal{B}} \in \mathbb{R}^q$ of the true mean $\bm{\psi}$ in the selected model, defined by $$(\bm{\beta}_{\mathcal{A}}^\top, \bm{b}_{\mathcal{B}}^\top)^\top = \breve{\bm{V}} \bm{\psi} :=  (\bm{C}^\top \bm{R}^{-1} \bm{C})^{-1} \bm{C}^\top \bm{R}^{-1} \bm{\psi}$$ with $\bm{C} = (\bm{X}|\bm{Z})$. 
To test some null hypothesis on selected fixed effects $H_0: \beta_{\mathcal{A}_j} = \beta_{j0}$ or  on selected random effects $H_0: b_{\mathcal{B}_j} = b_{j0}$, we use the conditional distribution $\bm{Y}|\,\bm{\psi}, \mathcal{S}(\bm{Y}) = (\mathcal{A}, \mathcal{B})$, $\beta_{\mathcal{A}_j} = \breve{\bm{v}}^\top \bm{\psi} := \bm{e}_j^\top \breve{\bm{V}} \bm{\psi} =
\bm{e}_j^\top (\bm{\beta}_{\mathcal{A}}^\top, \bm{b}_{\mathcal{B}}^\top)^\top $ the $j$th entry in $\beta_{\mathcal{A}_j}$  or analogously for $\bm{b}_{\mathcal{B}_j}$, and define the selection procedure $\mathcal{S}(\cdot)$ to return a tuple of set indices for fixed and random effects.
 Alternatively, we may want to test a group of coefficients or a certain linear combination 
\begin{equation}
H_0: \bm{C}_z \begin{pmatrix} \bm{\beta}_{\mathcal{A}} \\ \bm{b}_{\mathcal{B}} \end{pmatrix} = {\phi}_0
\end{equation}
for some row vector $\bm{C}_z \in \mathbb{R}^{1 \times (p+q)}$ and value ${\phi}_0 \in \mathbb{R}$. In this case, we denote $\breve{\bm{v}}^\top := \bm{C}_z \breve{\bm{V}}$. For simplicity, we state the proposal using $\breve{\bm{v}}^\top= \bm{e}_j^\top \breve{\bm{V}}$ in the following, but it can be equally applied for the more general linear combination case, replacing $\bm{e}_j^\top$ by $\bm{C}_z$.

\subsubsection{Null Distribution and Test Statistic} \label{condteststat}

We here relax the assumption of known $\bm{\Sigma}$ or known $\bm{Z}$ and  $\bm{G}$ and only assume that $\bm{R}$ is known. This allows us to conduct inference also in the case where the linear predictor or the random effect structure $\bm{Z}\bm{b}$ of the mixed model is misspecified, whereas in the marginal setup, the assumption of known $\bm{\Sigma}$ requires the corresponding random effect structure $\bm{Z}$ to be correct. Starting from a working linear mixed model with known covariance $\bm{R}$ of $\bm{\varepsilon}$ and (working) covariance $\bm{G}$  of the (working) random effects $\bm{b}$, 
the fixed and random effects can be predicted using $$\left( \begin{matrix} \hat{\bm{\beta}} \\ \hat{\bm{b}} \end{matrix} \right) = (\bm{C}^\top \bm{R}^{-1} \bm{C} + \bm{A})^{-1} \bm{C}^\top \bm{R}^{-1} \bm{Y} =: \bm{V} \bm{Y} \quad  \text{with } \bm{A} = \left( \begin{matrix} \bm{0} & \bm{0} \\ \bm{0} & \bm{G}^{-1} \end{matrix} \right).$$
In contrast to the linear model and marginal perspective in the previous subsection, we here distinguish between $\breve{\bm{v}}^\top$ defining the $j$th SDE $\breve{\bm{v}}^\top \bm{\psi}$ and the test vector $\bm{v}^\top = \bm{e}_j^\top \bm{V}$ (or $\bm{v}^\top = \bm{C}_z {\bm{V}}$)
that is used to test the null hypothesis $H_0: \breve{\bm{v}}^\top \bm{\psi} = \breve{\rho}$. 
The test statistic is $T = \bm{v}^\top \bm{Y} 
$ 
and  the conditional distribution of $T$ (without selection) is normal with $\mathbb{E}(T|\bm{\psi}) = \bm{v}^\top\bm{\psi} =: \rho$ and  
\begin{equation}
\text{Cov}(T | \bm{\psi}) = \bm{v}^\top \bm{R} \bm{v}= \bm{e}_j^\top (\bm{C}^\top \bm{R}^{-1} \bm{C} + \bm{A})^{-1} \bm{C}^\top \bm{R}^{-1} \bm{C} (\bm{C}^\top \bm{R}^{-1} \bm{C} + \bm{A})^{-1} \bm{e}_j . 
\label{CovVar1}
\end{equation}
For the null hypothesis $H_0: \breve{\bm{v}}^\top \bm{\psi} = \breve{\rho}$, the null distribution of $T$ in general is $\mathcal{N}(\rho, {\bm{v}}^\top \bm{R} {\bm{v}})$ with $\rho \neq \breve{\rho}$ if $\bm{G}\neq \bm{0}_{q \times q}$.
That is, the test statistic $T$ is not unbiased for the respective $\breve{\rho}$ due to the shrinkage induced by the covariance $\bm{G}$. 
Since $\bm{R}$ is assumed to be known for the moment (but cf.\ Section \ref{sec.limits}) and we use a known working covariance for $\bm{G}$ in $\bm{A}$, we can compute $T$ and derive its conditional distribution given $\bm{\psi}$ and the selection event. The choice of ${\bm{G}}$ affects the amount of shrinkage we use in the test statistic (and potentially the power of the test). 
In practice, we use the estimated covariance matrix $\hat{\bm{G}}$. If we set $\bm{G} = \bm{0}_{q \times q}$, the two vectors $\bm{v} \equiv \breve{\bm{v}}$ coincide. We compare both options ($\bm{G}= \hat{\bm{G}}$ vs. $\bm{G}=\bm{0}_{q \times q}$) in practice in Section~\ref{sec.sim}.

As (\ref{CovVar1}) does not account for the shrinkage bias 
in general, we suggest the use of a Bayesian covariance, which  yields better coverage when used in confidence intervals \citep{Nychka.1988, Marra.2012} and is thus also used in frequentist inference. 
We use the Bayesian covariance
\begin{equation}
\bm{K}^{-1} = (\bm{C}^\top \bm{R}^{-1} \bm{C} + \bm{A})^{-1} \label{bayesCov}
\end{equation}
and allow to replace (\ref{CovVar1}) by $\bm{e}_j^\top \bm{K}^{-1} \bm{e}_j$ in our inference framework. Alternatively, when using the covariance (\ref{CovVar1}), we rely on an asymptotic argument of the shrinkage effect decreasing with the sample size.
We investigate the efficacy and impact of our approach in the simulation section. 

 This approach allows to conduct inference in a similar manner as for the linear model. In particular, if the restriction of the space of $T$ induced by the selection procedure can be described as affine inequalities as in \citet{Loftus.2015a}, a corresponding truncated normal distribution for $T$ can be derived and inference can be conducted based on this distribution \citep[see, e.g.,][]{Ruegamer.2018b}. 
Given the working random effect covariance ${\bm{G}}$ and random effect structure $\bm{Z}$, the test vector $\bm{v}$ is fixed and we can produce samples in the direction of $\bm{v}$ in the same way as for the marginal perspective. We therefore define $\bm{P}_{\bm{v}} = \bm{R} \bm{v}\bm{v}^\top / (\bm{v}^\top \bm{R} \bm{v})$ and decompose $\bm{Y}$ into $\bm{P}_{\bm{v}} \bm{Y}$ in the direction of $\bm{v}$ and an orthogonal complement $\bm{P}_{\bm{v}}^\bot \bm{Y}$. We then generate samples  $\bm{Y}^b = \bm{R}\bm{v} / (\bm{v}^\top \bm{R} \bm{v}) \cdot T^b + \bm{P}_{\bm{v}}^\bot \bm{y}$ by drawing $T^b$ from a proposal distribution, checking for congruency with the original selection $\mathcal{S}(\bm{y}) = (\mathcal{A},\mathcal{B})$ and reweighting the samples using importance weights to approximate the expectation of formula \eqref{eq:pvalYang}. 

\indent

\subsection{Dealing with Unknown Error Covariance} \label{sec.limits}

As a known error covariance structure is not a realistic assumption in practice, \citet{Ruegamer.2018b} provided results on the effect when plugging in different estimates for the true variance in LMs. 
These findings, however, assume a diagonal error covariance matrix. We therefore extend their work and investigate conservative estimators as plug-in solutions for both proposed approaches in linear mixed models. 

For the marginal approach our proposal is to use $\hat{\bm{\Sigma}}_c$, a conservative estimator for $\bm{\Sigma}$, which assumes the random effect structure $\bm{Z}$ is fixed, i.e., not part of the model selection, and $\bm{G}$ as well as the error variances are unknown but estimated.  $\hat{\bm{\Sigma}}_c$ is given by the variance-covariance estimator of the intercept model with the given random effects structure, 
i.e., $\bm{Y} = \bm{1} \beta_0 + \bm{Z}\bm{b} + \bm{\varepsilon}$, which leaves as much variance in the response as possible unexplained by the mean model. 
This estimator for $\bm{\Sigma}$ is also used in the newly proposed test statistic $T_{\text{eff}}$, which depends on $\bm{\Sigma}$. An alternative approach in the case of grouped data with balanced designs, which we do not pursue here, could estimate $\bm{\Sigma}$ as a block-diagonal matrix with unstructured covariance blocks from an intercept-only model. For the conditional approach 
the random effects structure is part of the working model and we use the estimated random effects covariance $\hat{\bm{G}}$. We therefore can build on the results by \citet{Ruegamer.2018b} plugging in a variance estimator for the residual variance and investigate the effect in the simulation studies. When using the Bayesian covariance (\ref{bayesCov}) we replace $\bm{G}$ with the corresponding working covariance from the definition of $\bm{v}$.
 
\section{Additive Models} \label{sec.AM}

As additive models can be estimated using a mixed model representation, our proposed approach also provides ways to conduct selective inference for additive (mixed) models when using the conditional perspective. For illustrative purposes assume that the model selection results in a simple additive (working) model of the form $\bm{Y} = f(\bm{z}) + \bm{\varepsilon}$ with $\bm{\varepsilon} \sim \mathcal{N}(\bm{0},\bm{R})$ with diagonal matrix $\bm{R}$, covariate $\bm{z}$ and $f(\bm{z})$ denoting component-wise evaluation of $f$ on $\bm{z}$. Inference statements for more complex models with additional linear, non-linear or random terms can be derived in the same manner by extending the following design matrices. We assume $\bm{R} = \sigma^2 \bm{I}_n$ with $\bm{I}_n$ the $n$-dimensional identity matrix. The non-linear function $f$ can be approximated using a spline basis-function representation with $d$ basis functions $B_1, \ldots, B_d$ and for a given value $z$, this yields the design vector $\bm{C}_z = (B_1(z), \ldots, B_d(z))$ and spline approximation $f(z) \approx \bm{C}_z \bm{\gamma}$. The design matrix  $\bm{C}$ contains rows $\bm{C}_{z_i}, i=1, \dots, n$. The coefficients $\bm{\gamma}$ can be estimated using least squares or, alternatively, using  penalized least squares (PLS) 
$\min_{\bm{\gamma}} ||\bm{y}-\bm{C} \bm{\gamma}||_2^2 + \lambda \bm{\gamma}^\top \bm{P} \bm{\gamma}$ 
to induce smoothness of the estimated function $\hat{f}$ using some penalty matrix $\bm{P}$ and smoothing parameter $\lambda$. This is commonly done by re-parameterization $\bm{P}$ as $\text{diag}(\bm{0}_p, \bm{1}_{q})$ rewriting the PLS criterion as a linear mixed model \eqref{LMM} with $\bm{\gamma} = (\bm{\beta}^\top, \bm{b}^\top)^\top$, $\bm{G} = \tau^2 \bm{I}_q$, and $\tau^2  = \sigma^2 / \lambda$  \citep[see, e.g.,][]{Ruppert.2003}. 

Model selection in this case is more appropriate when considered from a conditional perspective, as $f(\bm{z})$ is represented in this approach using both fixed and random effects $\bm{X\beta} + \bm{Zb}$. The conditional distribution of $\bm{Y}|\bm{\psi}$ keeps $f$ fixed across observations in the mean structure and treats the random effects in the linear mixed model as just a mathematical tool (working model) to estimate $f$ with regularization 
\citep{Greven.2010}. We therefore  follow the conditional inference approach as discussed in Section \ref{cond} for additive models. An estimate of $f(z)$ for a certain value $z$ is given by $\hat{f}(z) = \bm{C}_z ( \hat{\bm{\beta}}^\top,  \hat{\bm{b}}^\top )^\top$. A test vector can thus be defined as 
\begin{equation}
\bm{v}_z^\top = \bm{C}_z (\bm{C}^\top \bm{R}^{-1} \bm{C} + \bm{A})^{-1} \bm{C}^\top \bm{R}^{-1}, \label{vInAM}
\end{equation}
with $\bm{A}$ defined as $\lambda \bm{P} = \sigma^2/\tau^{2} \bm{P}$.
Compared to Section~\ref{cond}, this replaces the $j$th unit vector $\bm{e}_j$ with the linear combination inducing vector $\bm{C}_z$. As before, we use $T_z = \bm{v}_z^\top \bm{Y}$ as our test statistic. In the case of $M$ non-linear functions $f_m$, $\bm{A}$ is defined as a block-diagonal matrix with blocks $\lambda_m \bm{P}_m, m=1,\ldots,M$, with smoothing parameter $\lambda_m$ and penalty matrix $\bm{P}_m$ corresponding to the $m$th term $f_m$.

For point-wise confidence bands for non-linear functions $f$, we investigate two possible options. The first idea  conditions on any  random effects in $\bm{\psi}$ and uses the conditional variance $\bm{v}_z^\top \bm{R} \bm{v}_z$ of the test statistic $T_z = \bm{v}_z^\top \bm{Y}$, analogous to the linear case. An alternative way to construct confidence intervals uses the Bayesian covariance matrix
\begin{equation}
 \bm{C}_z (\bm{C}^\top \bm{R}^{-1} \bm{C} + \sigma^2 /\tau^{2} \bm{P})^{-1} \bm{C}_z^\top \label{eq:BayesCov}
\end{equation}
to correct for the shrinkage effect, which leads to biased estimates even if the model is correct.
As for the linear (mixed) model, we use a Monte Carlo approximation for the p-value or confidence interval by generating samples $T^b_z$ from the null or proposal distribution.

\section{Simulation} \label{sec.sim}

We evaluate the proposed frameworks for linear mixed models and additive models in two simulation setups. For additive models we investigate the selective distribution of several fixed SDEs after selection via the conditional AIC \citep[see, e.g.,][]{Greven.2010}, whereas the selection criterion for the linear mixed model setup and either the conditional or marginal perspective is done with backward model selection based on significant tests using the package \texttt{lmerTest} \citep{lmerTest.2017}.

The purpose of this section is to investigate power properties of the proposed approaches -- a particular strength of our method -- as well as demonstrate the validity of the resulting inference. While for the following settings with standard selection procedures, selective inference results can be obtained within minutes or hours on a single machine, we refer the readers to the application section and corresponding appendix for more information on required run-times for very complex, potentially multi-step selection procedures with also a large number of Monte Carlo samples as required in our application.

\subsection{Mixed Model}

We first evaluate the proposed frameworks for linear mixed models, where the selection mechanism is a successive model reduction of fixed effects based on p-values using the package \texttt{lmerTest} \citep{lmerTest.2017}, keeping the random effects structure fixed (random effects selection will be a focus in \ref{ammsel}). The true data generating model is given by $Y_{ij} = \beta_0 + \sum_{k=1}^3 x_{k,ij} \beta_k + b_{0,j} + x_{2,ij} b_{1,j} +  \varepsilon_{ij}$ 
for $i = 1,\ldots,5$, $j=1,\ldots,30$, fixed effects $(\beta_0,\beta_1,\beta_2,\beta_3) = (1,2,-1,-2)$, random effects $b_{0,j} \sim \mathcal{N}(0,4)$, $b_{1,j} \sim \mathcal{N}(0,2)$, $cor(b_{0,j},b_{1,j}) = 0.5$, three additional noise variables $x_{4,ij},x_{5,ij},x_{6,ij}$ and residual $\varepsilon_{ij} \sim \mathcal{N}(0,\sigma^2)$, where $\sigma := \frac{\text{sd}(\bm{\eta})}{ \text{SNR}}$ is defined via the signal-to-noise ration $\text{SNR} \in \{2,4 \}$ with $\bm{\eta}$ the linear predictor vector of all $n = 150$ observations in the data generating process. All covariates are drawn independently from a standard normal distribution, yielding a maximum empirical correlation of around 0.17. To investigate the impact of using the true (co-)variance $\Sigma$ or a plug-in estimator, we compare the true (co)variance (``Truth''), the estimated (co-)variance of the chosen model (``Model Estimate''), the estimated residual variance in the marginal covariance of the corresponding intercept model with fixed random effect structure (``ICM'') and the asymptotically conservative estimate $\text{var}(\bm{y})$ by \citet{Tibshirani.2015} (``Var(Y)''). As we keep fixed the random effects in our first simulation, we also investigate the ICM plug-in for the conditional perspective, where we use a model with linear predictor $\beta_0 + b_{0,j} + x_{2,ij} b_{1,j}$. Note that the derivation by \citet{Tibshirani.2015} for the Var(Y) plug-in assumes $\bm{\Sigma} = \sigma^2 \bm{I}$, which in our simulations only holds for the conditional perspective. 
We use as many simulation iterations as necessary to obtain at least $100$ data sets with selection of the correct or a supmodel. We base p-values for each data set on $B=500$ importance samples.

In Figure~\ref{fig:plot_sim_1} the observed pooled p-values for all noise variables (top row) and the two signal variables $x_1, x_2$ (bottom rows) for both conditional and marginal perspective (columns) are plotted against expected quantiles of the standard uniform distribution and compared to the naive p-values. 
Results indicate that 
naive p-values 
exhibit non-uniformity under the null. By contrast, selective p-values show similarly high power for the signal variables for all settings and uniformity under the null.
The comparison of variance estimates indicates that while the true variance-covariance matrix yields well-calibrated inference, all estimates yield conservative inference. The inference for the model estimate is similar to that for the truth or even conservative here, while the conservative estimates ``ICM'' and ``Var(Y)'' are even more conservative. This finding encourages the use of the model estimate as plugin for the true variance. We also observe that there is no notable difference in using the marginal or conditional perspective in the given settings.

\begin{knitrout}
\definecolor{shadecolor}{rgb}{0.969, 0.969, 0.969}\color{fgcolor}\begin{figure}[ht]
\includegraphics[width=\maxwidth]{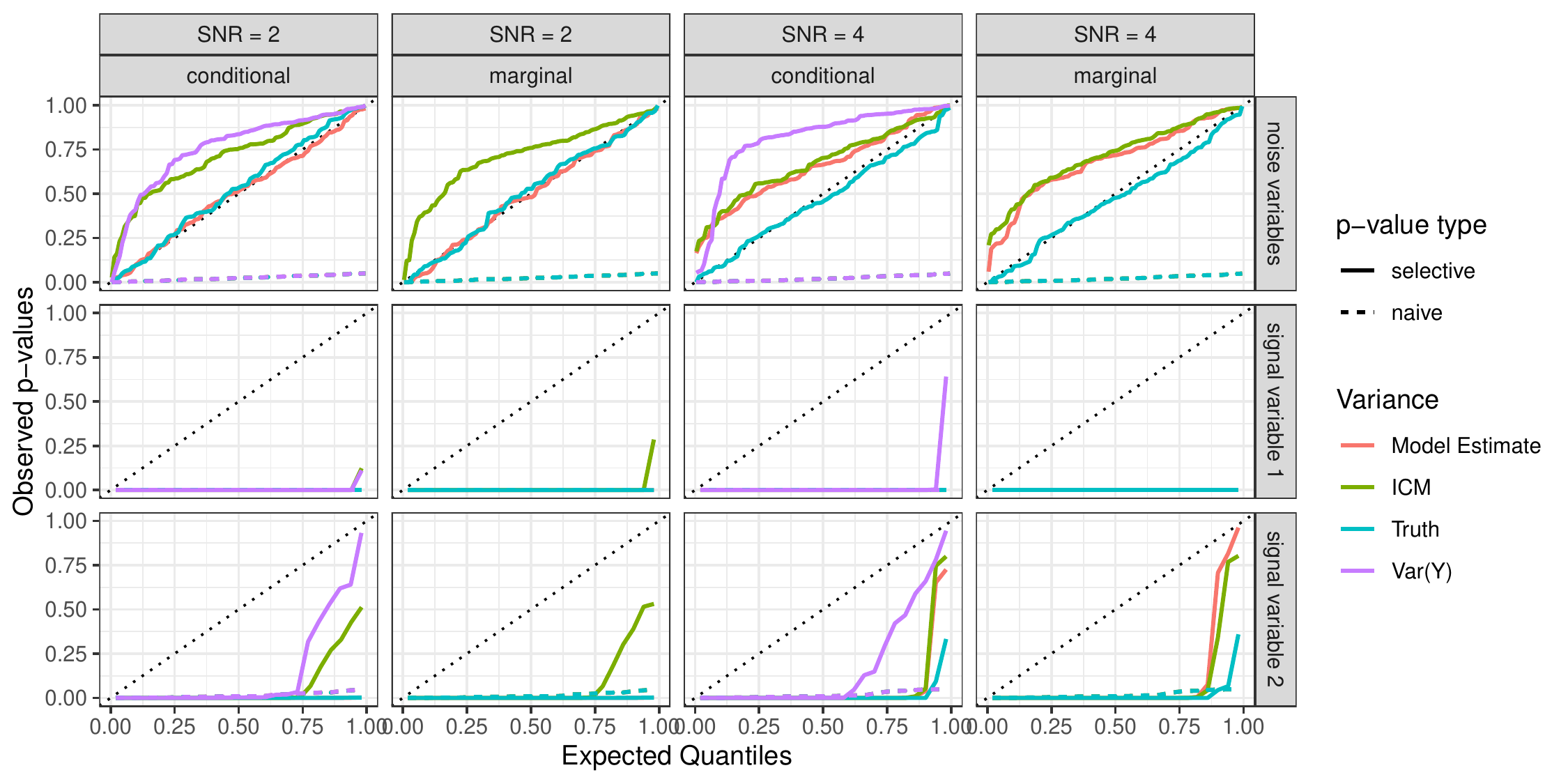} \caption[Quantiles of the standard uniform distribution versus the observed p-values for noise variables (top row) as well as two signal variables (bottom rows), each for different SNR as well as for both mixed model perspectives (columns) and different variances used in calculating the p-values (colours)]{Quantiles of the standard uniform distribution versus the observed p-values for noise variables (top row, pooled across noise variables $x_4$ to $x_6$) as well as two signal variables $x_1$ and $x_2$ (bottom rows), each for different SNR as well as for both mixed model perspectives (columns) and different variance estimates used in calculating the p-values (colours). To ensure that the null hypothesis is correct for the noise variables, data points in this plot are based on iterations where the true model was selected or a model in which the true model is nested. Solid lines represent selective p-values, dashed lines represent the corresponding naive p-values not accounting for selection.}\label{fig:plot_sim_1}
\end{figure}

\end{knitrout}

For the same simulation setting, we additionally investigate the influence of the choice of matrix $\bm{G}$ for the conditional perspective discussed in Section~\ref{condteststat}. We use the same data generating process using the model estimate plugin for the variance and $\text{SNR} = 4$. We compare the proposed working covariance against a zero matrix to assess the influence of shrinkage on the selective p-values. The corresponding results are given in the Supplementary Material~\ref{sec.sim_extra}. Results show 1) uniformity under the Null for both versions with little difference in their distribution for both signal and noise variables, 2) no notable difference in power between the two approaches when reducing the covariate effects to $\beta_1 = 0.5 = -\beta_3$ and $\beta_2=0.25$ with smaller $\text{SNR} = 2$.

\subsection{Additive Model Selection}\label{ammsel}

We also evaluate our approach for additive models using the cAIC as selection criterion to select among different additive regression models. The true data generating model for $n = 500$ observations is given by $y_i = 1 + f_1(z_{i,1}) + f_2(z_{i,2}) + \varepsilon_i,$ where $f_1(z) = -\tanh(z)$, $f_2(z) = \sin(3z)$, $\varepsilon_i \sim \mathcal{N}(0,\sigma^2)$ for $i=1,\ldots,n$ and $\sigma^2 \in \{1,10\}$. $n/2$ observations of the two signal variables as well as of the two further noise variables $z_3, z_4$ are independently drawn from a standard normal distribution. We then add corresponding $n/2$  observations with minus these $z_1$ to $z_4$ values to  fulfill the condition $\sum_i f_j(z_{i,j}) = 0, j=1,\dots, 4$ by construction. We thereby ensure that covariates with non-linear effects are not affected by sum-to-zero constraints and thus fix the locations where the functions $f_1, f_2$ cross zero. We select among five different regression models, where either all covariates are assumed to have a linear effect, only $z_1$, only $z_3$, $z_1$ and $z_2$, or $z_1$ and $z_3$ are estimated as having a non-linear effect. In each of the 500 simulation iterations, these five different models are compared using the cAIC. For selected covariates chosen as having a non-linear effect, point-wise p-values for $H_0: f_j(\upsilon) = 0$ are calculated for two specific values $\upsilon \in \{-1,0\}$. This is done for $j=1$ and $j=2$ where the null hypothesis is true for $\upsilon = 0$ but does not hold for $\upsilon = 1$. If the selected covariate $z$ is modeled as having a linear effect $z_j\beta_j$, the corresponding effect $\beta_j$ is tested against zero. This is done for $j=3$ and $j=4$, for which the null hypothesis $\beta_j = 0$ is true. To investigate the impact of using different variances for sampling, we compare the usage of the true variance (``True''), the estimated variance of the chosen model (``Model Estimate'') and the conservative estimate $\text{var}(\bm{y})$ (``Var(Y)''). For all combinations, we also investigate the difference between using the Bayesian covariance as in (\ref{eq:BayesCov}) and the classical covariance. 

In Figure~\ref{fig:plot_sim_2} the observed p-values for the two noise variables combined (first row) and the signal variables at the two pre-defined locations (last four rows) are plotted against expected quantiles of the standard uniform distribution. Results indicate that all settings for the true variance or estimated variance and non-Bayesian covariance definition reveal uniform p-values under the null for both noise variables and locations $f_1(0), f_2(0)$ where the non-linear functions cross zero. The difference in power due to $|f_1(-1)| > |f_2(-1)|$ becomes apparent when comparing the rows two and four. Using the Bayesian covariance definition, p-values are uniform for the noise variables and tend to be conservative where $f_1$ or $f_2$ are truly zero, while not yielding larger power for the estimated or true variance. These results encourage the use of the classical covariance definition as well as the estimated variance as plugin estimator. When using the conservative estimator Var(Y), results are only notably affected for the classical covariance definition, yielding more conservative results in all simulation settings.

\begin{knitrout}
\definecolor{shadecolor}{rgb}{0.969, 0.969, 0.969}\color{fgcolor}\begin{figure}[ht]
\includegraphics[width=\maxwidth]{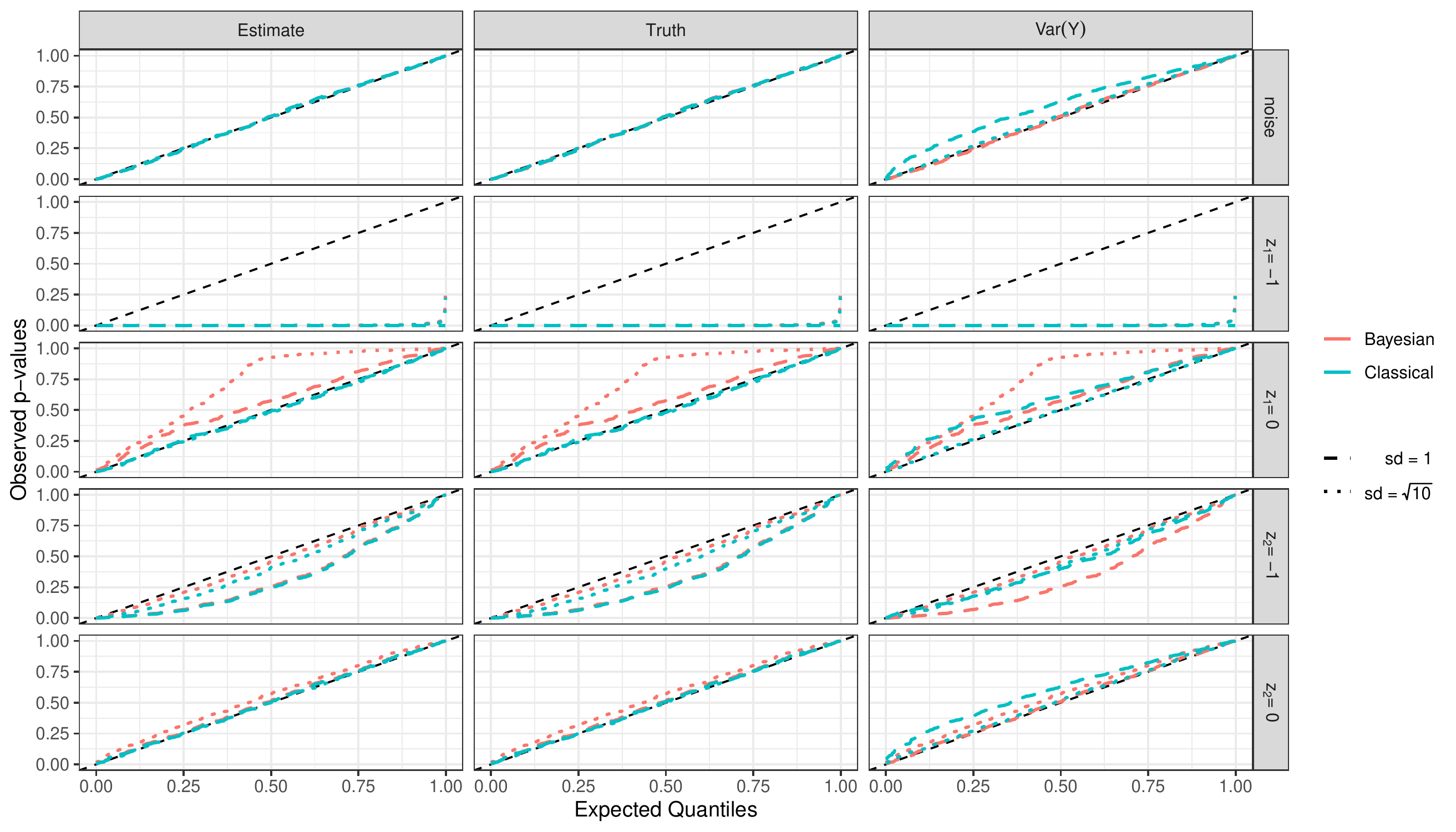} \caption[Quantiles of the standard uniform distribution versus the observed p-values for noise variables (top row) as well as the two locations for the two signal variables (bottom four rows), each for different error standard deviations (sd, different linetypes) and either the Bayesian or classical covariance definition (colour)]{Quantiles of the standard uniform distribution versus the observed p-values for noise variables (top row, pooled for $x_3$ and $x_4$) as well as the two locations for the two signal variables (bottom four rows), each for different error standard deviations (sd, different linetypes) and either the Bayesian or classical covariance definition (colours). The usage of different variance estimates for the error variance is visualized in different columns.}\label{fig:plot_sim_2}
\end{figure}

\end{knitrout}

\section{Application to the Determinants of Inflation} \label{sec.appl}

We now apply our proposed framework to the field of monetary economics to analyse the determinants of inflation in a large country sample. 

\subsection{Introduction}

The analysis of country-specific worldwide inflation rates (defined as the percentage changes in the consumer price index) has been subject to many empirical investigations \citep[see, e.g.,][]{RePEc:eee:moneco:v:52:y:2005:i:3:p:529-554, RePEc:chb:bcchwp:491}. Low and stable inflation rates are by now the established (main) goal of monetary policymakers around the world. In order to achieve this objective, a good understanding of the underlying inflation process is crucial for the effectiveness and efficiency of monetary policy. Economic theory proposes a variety of potential drivers of inflation. However, the question remains which economic theory and which corresponding variables provide the most persuasive answer to this question from an empirical point of view.  This question has been addressed by \citet{BRV_Inflation} utilizing an additive mixed model (AMM) approach with an extensive two-stage model-selection procedure based on the cAIC. \citet{BRV_Inflation} identify eight economic theories that are discussed in monetary economics as explanations of inflation (cf. Supplementary Material \ref{sec.AppendixDetails} for a short overview). It is not possible to assign the corresponding variables for each of these theories unambiguously as some uncertainty remains about the empirical variables that best approximate the variables motivated by economic theory. For example, the first theory comprises the \textit{Output Gap (\%)} variable and the \textit{Real GDP Growth (\%)} variable, (cf. Table \ref{Tab:Allocation1} in the Supplementary Material \ref{sec.AppendixTables}), where economic theory does not give a clear answer on which of these two variables represents the theory best from an empirical point of view. Consequently, a choice of different sets of empirical variables is assigned to each economic theory. These compilations of variables in the various AMMs are purely based on economic theory.  
The model selection procedure of \citet{BRV_Inflation} described in Section \ref{sec:AppSelection} is rather atypical due to missing values of certain predictors, which prohibits a direct comparison of all AMMs at the same stage. The resulting use of different subsets of the data in the course of a hierarchical selection procedure makes it infeasible to provide an analytic form for the selection condition restricting the inference space of $T$. Since the selection procedure is repeatable in a boostrap-like manner, our proposed Monte Carlo based framework, however, can handle this unusual selection procedure and is thus particularly suitable to derive inference statements for \citet{BRV_Inflation}.

\subsection{Data and Model Setup}

The setup of \citet{BRV_Inflation} is as follows: 27 (metric and categorical) predictors and the dependent variable \textit{inflation} (in percent), denoted by $\Tilde{y}_{i,t}$, are given for $i = 1,\ldots, n=124$ countries and for $t = 0,\ldots, T=18$ consequent years from 1997 to 2015 such that $\bm{\Tilde{y}_{i}} = (\Tilde{y}_{i,0},\ldots, \Tilde{y}_{i,T})^\top $. The vector $\bm{\Tilde{y}} = (\bm{\Tilde{y}_{1}}^\top,\ldots,\bm{\Tilde{y}_{n}^\top)^\top}$ has been transformed with the natural logarithm $\bm{y} := \ln(\bm{\Tilde{y}} + 10.86)$ after shifting the support to values $ \geq 1$ to avoid numerical instabilities. Due to missing information for some variables, $2.8\%$ of the data is generated by imputation. \citet{BRV_Inflation} computed their results on the first of five imputations due to the lacking theoretical underpinning for averaging random effect predictions across multiple imputations, but checked for robustness and found stable results with respect to the selection event for all five imputations. Future research might strive to incorporate imputation uncertainty into selective inference statements.

The generic AMM used to explain $y_{i,t}$ by a set of predictors $A_{j,l}$ is given in Equation \eqref{eq:AMM}. Each of the eight economic theories is represented by a set $G_l:=\{\{A_{1,l}\},\{A_{2,l}\},\ldots,\{A_{m_{l},l}\}\}, l = 1,\ldots,8$, containing $m_l:=\ \mid G_l\mid$ sets of predictors $A_{j,l}$. Each $A_{j,l}$ is composed of disjunct subsets $B_{j,l}$ and $C_{j,l}$ of predictors with linear and non-linear effects, respectively, as well as pairs $D_{j,l}$ of variables in $B_{j,l}$ and pairs $E_{j,l}$ of variables in $C_{j,l}$ with linear and non-linear interaction effects, respectively. Non-linear effects $h$ of predictors $x \in C_{j,l}$ are estimated by univariate cubic P-Splines \citep{PSplines} with second-order difference penalties. Interaction effects $f(\cdot,\cdot)$ of pairs $(x, x^{*})$ of variables in $E_{j,l}$ are modeled using penalized bivariate tensor-product splines. The assignment to $B_{j,l}$, $C_{j,l}$, $D_{j,l}$ and $E_{j,l}$ can be found in Tables \ref{Tab:Allocation1} and \ref{Tab:Allocation2} in the Supplementary Material \ref{sec.AppendixTables}. Each model $M_{j,l}$ corresponding to one $A_{j,l} \in G_l$ is of the following form:
\begin{equation}\label{eq:AMM}
y_{i,t} =  \beta_{0} 
+ \eta_{i,t}
+  \bm{Z_{i,t}b_{i}} + \epsilon_{i,t},
\end{equation}
\begin{equation*}
\eta_{i,t} = \sum_{x \in  B_{j,l}} x_{i,t} \beta_{x} 
+ \sum_{(x,x^{*}) \in D_{j,l}}  (x_{i,t}x_{i,t}^{*}) \beta_{(x,x^{*})}  
+ \sum_{x \in C_{j,l}} h_x(x_{i,t}) 
+ \sum_{(x,x^{*}) \in E_{j,l}}  f_{(x,x^{*})}(x_{i,t},x_{i,t}^{*})  
\end{equation*}
with $\bm{b_{i}} = (b_{i,0},b_{i,1})^\top \stackrel{iid}{\sim} N(\bm{0},\bm{G})$, where a random intercept $b_{i,0}$ and a random slope $b_{i,1}$ with design vector $\bm{Z_{i,t}}\equiv \bm{Z_{t}} = (1,t)$ and non-diagonal covariance $\bm{G}$ are (always) included to capture the serial within-country correlation. Further, $\bm{\epsilon_{i}} \sim N(\bm{0},\bm{R_{i}})$ is assumed with $\bm{\epsilon_{i}} \perp \!\!\! \perp  \bm{b_{i}}$, where $\bm{R_{i}}$ is a diagonal matrix with potentially heterogeneous country-specific variances $\sigma_{i}^{2}$ on its diagonal. 
In total, there are 90 (= $\sum_{l = 1}^{8} m_{l}$) such models for all predictor sets $A_{j,l}$ comprised by each economic theory $G_l$. For each $G_l$ there is one set of models $\mathcal{M}_l$ which includes all corresponding $M_{j,l}$.

\subsection{Model Selection Procedure}\label{sec:AppSelection}

The model selection procedure of \citet{BRV_Inflation} is as follows: at a first stage $\mathcal{S}_{fir}$, a winner model $M^{*}_{l}$ with the lowest cAIC among models $M_{j,l}$ in the set $\mathcal{M}_l$ is selected per theory. At a second-stage $\mathcal{S}_{sec}$, $M^{*}_{l}$, $l = 1,\ldots, 8$, are collected in the set $\mathcal{M}_{P}$. Some predictors associated with $\mathcal{M}_2$, $\mathcal{M}_3$ and $\mathcal{M}_4$ are not imputed as these predictors are restricted in availability either across time and/or countries which makes a direct model comparison by means of the Likelihood and thus the cAIC inadmissible. As a result, if the predictor sets included in $M^{*}_{2}$, $M^{*}_{3}$ and $M^{*}_{4}$ are only available for a subsample of data, they are instead added to $\mathcal{M}^{''}$ to be compared to the AMM with the lowest cAIC in $\mathcal{M}_{P}$ later. 
The winner $M_P$ has the lowest cAIC in the set of models $\mathcal{M}_{P}$ and its cAIC is finally compared to each $M^{''}\in \mathcal{M}^{''}$ on the corresponding different data subsets to yield the overall winner $M^{**}$. If the computation of any AMM on any subset of the data fails, this AMM is assigned the highest cAIC in the given comparison. This can happen in particular for complex models on smaller subsets of the data. First- and second-stage selection are together labeled $\mathcal{S}_{sec}$. $M^{**}$ represents the model with the highest empirical relevance for the application.  

Based on the described model selection procedure \citet{BRV_Inflation} obtain among other results, $M^{*}_{4} = M_{14,4}$, $M^{*}_{5} = M_{7,5}$ and $M^{**} = M_{9,7}$. Note that $M^{*}_{4}$ is estimated on a subsample of 80 countries and $M^{*}_{5}$ and $M^{**}$ on the full sample of 124 countries. We now provide inference statements for partial effect estimates within $M^{*}_{4}$, $M^{*}_{5}$ and $M^{**}$ taking into account the model selection uncertainty from the first- and second-stage.   Inference statements for $M^{**}$ condition on model selection in both stages, i.e.\ on $\mathcal{S}_{sec}$, to account for uncertainty in the empirical variables capturing each economic theory best. 

\subsection{Results}

\begin{figure}[htb]

    \centering
    \includegraphics[width=\textwidth]{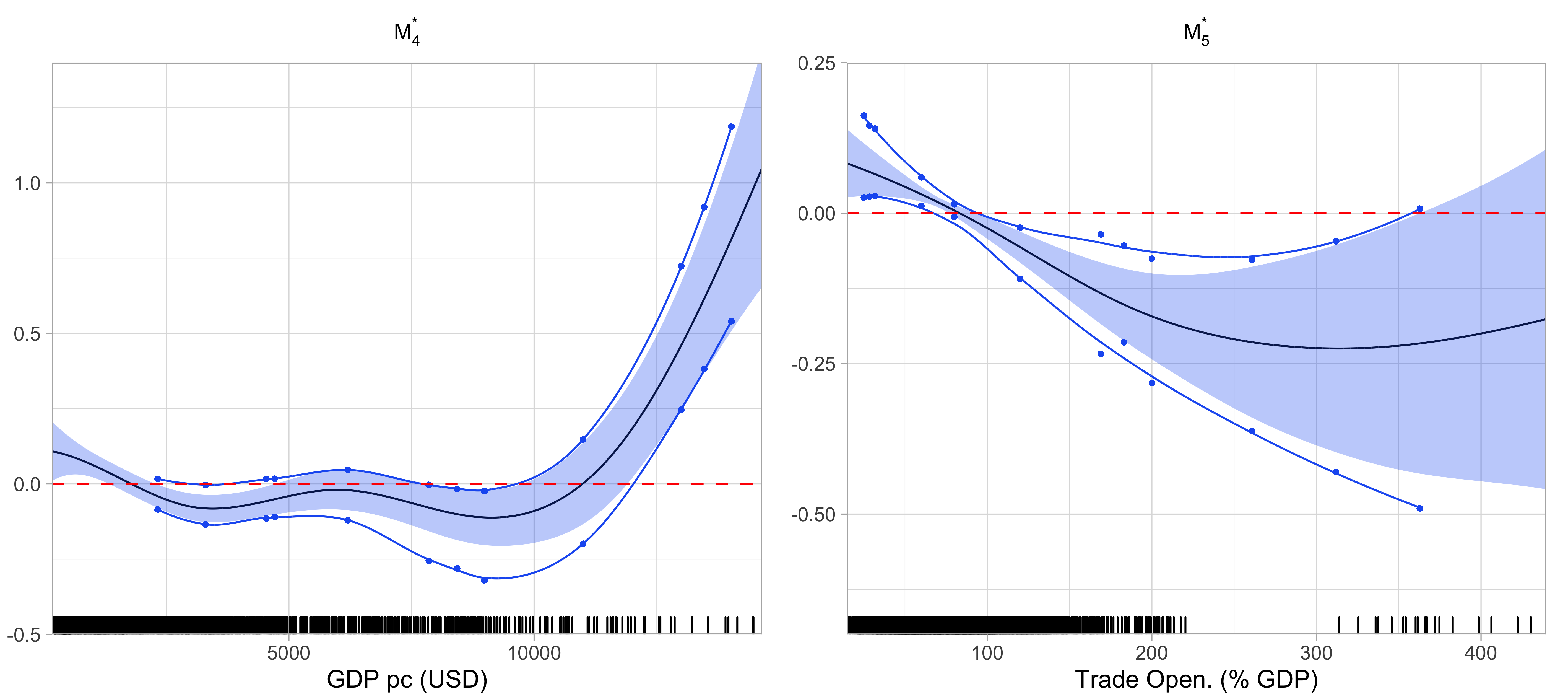}
    \caption[]{The left panel shows the partial effect estimate $\hat{f}_{GDPpc}$ included in $M^{*}_{4}$, and the right panel  the partial effect estimate $\hat{f}_{Open}$ included in $M^{*}_{5}$. The standard pointwise 95\%-Bayesian confidence intervals are shaded and the pointwise selective 95\%-Bayesian confidence intervals are shown as dots, connected using a LOESS estimate indicated by solid lines. 
    }
    \label{fig:ApplicationPoSI}

\end{figure}

\paragraph{Results for model $M^{*}_{4}$.} We pick 12 approximately equally spaced locations of interest $z^{+}\in\{2324.2, 3300.4,  
\ldots, 13470.4, 14023.6\}$ on the estimated spline function $\hat{f}_{GDPpc}$, which represents the partial effect estimate of the \textit{Real GDP per capita (USD)} variable, labeled \textit{GDP pc (USD)}, onto log-inflation. We provide 95\% (point-wise) confidence intervals for each projection $f_{GDPpc}(z^{+})$ based on our proposed selective inference procedure 
with $\alpha = 0.05$. 
We compute the corresponding test vector,  
draw $b = 1,\ldots,B \equiv 1500$ test statistics 
from a mixture of $4$ Gaussian distributions as proposal distribution in the importance sampling and compute inference results 
as described in Section \ref{MC}. 
The 
mixture distribution with different expectations as proposal distribution is based on empirical observations, yielding a substantial larger amount of usable samples in this application. 
\begin{table}[htb]
\centering
{\footnotesize
\begin{tabular}{l|c|c|c|c|c|c}
\toprule
$z^{+}$  & 2324.2 & 3300.4 & 4538.7 & 4711.0 & 6199.9 & 7852.4 \\
\midrule
\rowcolor{Gray}
  $0 \cdot t_{z^{+}}$ & 11\% & 2\% & 7\% & 14\% & 26\% & 27\% \\  
  $1/3 \cdot t_{z^{+}}$  & 18\% & 6\% & 20\% & 22\% & 26\% & 26\% \\  
  \rowcolor{Gray}
  $2/3 \cdot t_{z^{+}}$& 31\% & 30\% & 32\% & 30\% & 24\% & 25\% \\ 
  $3/3 \cdot t_{z^{+}}$ & 41\% & 62\% & 41\% & 34\% & 23\% & 22\% \\ 
  \midrule
  Total & 759 & 478 & 823 & 1027 & 1345 & 1386 \\   
  \midrule
  \midrule
  p-value & 0.184 & 0.034 & 0.118 & 0.128 & 0.438 & 0.043 \\ 
\bottomrule
\end{tabular}

}
\caption{Share of $1500$ samples that have led to the original selection for the first six $z^{+}$, separated by each component of the Gaussian mixture distribution (leftmost column), with mean values being a ratio of the observed test statistic. The percentages add up to 100\% which is equal to \textit{Total}. The last row contains the p-values that result from testing the null hypothesis. The results for all twelve $z^{+}$ can be found in the Supplementary Material.}
\label{Tab:Location4}
\end{table}
For the partial effect estimate $\hat{f}_{GDPpc}$ in $M^{*}_{4}$, the standard 95\%-Bayesian confidence intervals and the (in part wider) selective confidence intervals for each projection ${f}_{GDPpc}(z^{+})$ are shown in the left panel of Figure \ref{fig:ApplicationPoSI}. Corresponding p-values are given in the last row of Table \ref{Tab:Location4} for the first six locations. The complete table with all 12 locations and details on the run-time for the computation are given in the Supplementary Material. 

\paragraph{Results for model  $M_{5}^{*}$.} We pick 12 locations $z^{++} \in \{25.0, 28.4, 
\ldots, 311.9, 362.8\}$ on the estimated spline function $\hat{f}_{Open}$, which represents the estimated partial effect of the variable \textit{Trade Openness (\% of GDP)}, labeled \textit{Trade Open. (\% GDP)}, on log-inflation. We compute selective 
95\% (point-wise) confidence intervals for each projection ${f}_{Open}(z^{++})$ 
using importance sampling analogous to those discussed for $M_{4}^{*}$ above (but using $B^{++} = 1200$ for computational reasons). Results are visualized in the right panel of Figure \ref{fig:ApplicationPoSI}. Selective pointwise tests for $H_0: f_{Open}(z^{++}) = 0$ were significant at $\alpha = 0.05$ at all tested locations $z^{++}$  besides 80 and 362.8, even after adjusting for model selection (cf.\  Table \ref{Tab:Location5} in the Supplementary Material \ref{sec.AppendixTables}).
 

\paragraph{Results for model $M^{**}$.} Effects of four variables are modelled through three separate model terms which are included in the selected $A_{9,7}$ (cf. Table \ref{Tab:Allocation2} in the Supplementary Material \ref{sec.AppendixTables}). Inference statements for the linear effect of \textit{Real GDP Growth (\%)} and the non-linear univariate effect of \textit{Credit (\% of GDP) Growth} can be obtained in the same manner as done for $M_4^*$ and $M_5^*$. We here focus on the non-linear interaction effect of the two remaining variables, \textit{Energy Prices (USD)} and \textit{Energy Rents (\% of GDP)}, modelled using a penalized bivariate tensor-product spline. We conduct an overall test for the interaction surface, that is $$H_0: f(\textit{Energy Prices (USD), Energy Rents (\% of GDP)}) = 0$$ for all pairs (\textit{Energy Prices (USD), Energy Rents (\% of GDP)}). 
We therefore use the selective $\chi$-significance test for groups of variables, which is  applied to the group of spline coefficients. We use importance sampling analogous to those discussed for $M_{4}^{*}$ and $M_{5}^{*}$ using $B^{*} = 4000$. Based on 80 samples that are consistent with the original model selection, we compute p-values. These suggest that the null hypothesis can be rejected at the pre-defined significance level (cf. Supplementary Material \ref{sec.AppendixDetails} for details). 

\section{Discussion} \label{sec.disc}

In this work we discuss extensions of selective inference to linear mixed models and additive (mixed) models. We establish new test statistics, hypotheses of interest and working model assumptions and introduce a conditional perspective for this working model. 
We show how recent proposals for selective inference in linear models can be transferred to these larger model classes and provide evidence of the validity of our approach using simulation studies. 
The application of the proposed approach to the determinants of inflation underlines the usefulness of our approach by allowing to compute p-values and confidence intervals for additive models with non-i.i.d.\ errors after a non-trivial multi-stage selection procedure including different model types, missing values and varying numbers of observations for the different model comparisons. 
The proposed Monte Carlo approximation allows for arbitrary selection mechanisms $\mathcal{S}$ as long as $\mathcal{S}$ is deterministic on a given data set, 
yielding statistically valid inference. 
The approximation is, however, computationally expensive. In particular for additive models, where selective confidence intervals for non-linear functions require $B$ reruns of $\mathcal{S}$ for each point on the function that is to be tested, future research is needed to circumvent this computational bottleneck. 

\clearpage

\begin{small}

\spacingset{1}

\bibliographystyle{chicago}
\bibliography{bibliography.bib}

\begin{thebibliography}{}

\bibitem[\protect\citeauthoryear{Baumann, Rossi, and Volkmann}{Baumann
  et~al.}{2020}]{BRV_Inflation}
Baumann, P., E.~Rossi, and A.~Volkmann (2020).
\newblock {What Drives Inflation and How: Evidence from Additive Mixed Models
  Selected by cAIC}.
\newblock {\em arXiv e-prints arXiv:2006.06274\/}.

\bibitem[\protect\citeauthoryear{Berk, Brown, Buja, Zhang, Zhao, et~al.}{Berk
  et~al.}{2013}]{Berk.2013}
Berk, R., L.~Brown, A.~Buja, K.~Zhang, L.~Zhao, et~al. (2013).
\newblock Valid post-selection inference.
\newblock {\em The Annals of Statistics\/}~{\em 41\/}(2), 802--837.

\bibitem[\protect\citeauthoryear{Calder\'{o}n and Hebbel}{Calder\'{o}n and
  Hebbel}{2008}]{RePEc:chb:bcchwp:491}
Calder\'{o}n, C. and K.~S. Hebbel (2008, Oct).
\newblock What drives inflation in the world?
\newblock Working Papers Central Bank of Chile 491, Central Bank of Chile.

\bibitem[\protect\citeauthoryear{Cat\~{a}o and Terrones}{Cat\~{a}o and
  Terrones}{2005}]{RePEc:eee:moneco:v:52:y:2005:i:3:p:529-554}
Cat\~{a}o, L.~A. and M.~E. Terrones (2005, April).
\newblock Fiscal deficits and inflation.
\newblock {\em Journal of Monetary Economics\/}~{\em 52\/}(3), 529--554.

\bibitem[\protect\citeauthoryear{Eilers and Marx}{Eilers and
  Marx}{1996}]{PSplines}
Eilers, P.~H. and B.~D. Marx (1996).
\newblock Flexible smoothing with b-splines and penalties.
\newblock {\em Statistical science\/}, 89--102.

\bibitem[\protect\citeauthoryear{{Fithian}, {Sun}, and {Taylor}}{{Fithian}
  et~al.}{2014}]{Fithian.2014}
{Fithian}, W., D.~{Sun}, and J.~{Taylor} (2014).
\newblock {Optimal Inference After Model Selection}.
\newblock {\em arXiv e-prints arXiv:1410.2597\/}.

\bibitem[\protect\citeauthoryear{Greven and Kneib}{Greven and
  Kneib}{2010}]{Greven.2010}
Greven, S. and T.~Kneib (2010).
\newblock On the behaviour of marginal and conditional {AIC} in linear mixed
  models.
\newblock {\em Biometrika\/}~{\em 97\/}(4), 773--789.

\bibitem[\protect\citeauthoryear{Kuznetsova, Brockhoff, and
  Christensen}{Kuznetsova et~al.}{2017}]{lmerTest.2017}
Kuznetsova, A., P.~B. Brockhoff, and R.~H.~B. Christensen (2017).
\newblock {lmerTest} package: Tests in linear mixed effects models.
\newblock {\em Journal of Statistical Software\/}~{\em 82\/}(13), 1--26.

\bibitem[\protect\citeauthoryear{Lee, Sun, Sun, and Taylor}{Lee
  et~al.}{2016}]{Lee.2016}
Lee, J.~D., D.~L. Sun, Y.~Sun, and J.~E. Taylor (2016, 06).
\newblock Exact post-selection inference, with application to the lasso.
\newblock {\em The Annals of Statistics\/}~{\em 44\/}(3), 907--927.

\bibitem[\protect\citeauthoryear{{Loftus} and {Taylor}}{{Loftus} and
  {Taylor}}{2015}]{Loftus.2015a}
{Loftus}, J.~R. and J.~E. {Taylor} (2015).
\newblock {Selective inference in regression models with groups of variables}.
\newblock {\em arXiv e-prints arXiv:1511.01478\/}.

\bibitem[\protect\citeauthoryear{Marra and Wood}{Marra and
  Wood}{2012}]{Marra.2012}
Marra, G. and S.~N. Wood (2012).
\newblock Coverage properties of confidence intervals for generalized additive
  model components.
\newblock {\em Scandinavian Journal of Statistics\/}~{\em 39\/}(1), 53--74.

\bibitem[\protect\citeauthoryear{Nychka}{Nychka}{1988}]{Nychka.1988}
Nychka, D. (1988).
\newblock Bayesian confidence intervals for smoothing splines.
\newblock {\em Journal of the American Statistical Association\/}~{\em
  83\/}(404), 1134--1143.

\bibitem[\protect\citeauthoryear{Overholser and Xu}{Overholser and
  Xu}{2014}]{overholser2014effective}
Overholser, R. and R.~Xu (2014).
\newblock Effective degrees of freedom and its application to conditional aic
  for linear mixed-effects models with correlated error structures.
\newblock {\em Journal of multivariate analysis\/}~{\em 132}, 160--170.

\bibitem[\protect\citeauthoryear{R{\"u}gamer and Greven}{R{\"u}gamer and
  Greven}{2018}]{Ruegamer.2018b}
R{\"u}gamer, D. and S.~Greven (2018).
\newblock Selective inference after likelihood- or test-based model selection
  in linear models.
\newblock {\em Statistics \& Probability Letters\/}~{\em 140}, 7 -- 12.

\bibitem[\protect\citeauthoryear{R{\"u}gamer and Greven}{R{\"u}gamer and
  Greven}{2020}]{Ruegamer.2018c}
R{\"u}gamer, D. and S.~Greven (2020).
\newblock {Inference for L2-Boosting}.
\newblock {\em Statistics and Computing\/}~{\em 30\/}(2), 279--289.

\bibitem[\protect\citeauthoryear{Ruppert, Wand, and Carroll}{Ruppert
  et~al.}{2003}]{Ruppert.2003}
Ruppert, D., M.~P. Wand, and R.~J. Carroll (2003).
\newblock {\em Semiparametric regression}.
\newblock Cambridge series in statistical and probabilistic mathematics.
  Cambridge and New York: Cambridge University Press.

\bibitem[\protect\citeauthoryear{{S{\"a}fken}, {Kneib}, {van Waveren}, and
  {Greven}}{{S{\"a}fken} et~al.}{2014}]{Saefken.2014}
{S{\"a}fken}, B., T.~{Kneib}, C.~{van Waveren}, and S.~{Greven} (2014).
\newblock A unifying approach to the estimation of the conditional akaike
  information in generalized linear mixed models.
\newblock {\em Electronic Journal of Statistics\/}~{\em 8\/}(1), 201--225.

\bibitem[\protect\citeauthoryear{{S{\"a}fken}, {R{\"u}gamer}, {Kneib}, and
  {Greven}}{{S{\"a}fken} et~al.}{2019}]{Saefken.2019}
{S{\"a}fken}, B., D.~{R{\"u}gamer}, T.~{Kneib}, and S.~{Greven} (2019).
\newblock {Conditional Model Selection in Mixed-Effects Models with cAIC4}.
\newblock {\em Journal of Statistical Software\/}.
\newblock to appear.

\bibitem[\protect\citeauthoryear{Tibshirani, Rinaldo, Tibshirani, and
  Wasserman}{Tibshirani et~al.}{2018}]{Tibshirani.2015}
Tibshirani, R.~J., A.~Rinaldo, R.~Tibshirani, and L.~Wasserman (2018, 06).
\newblock Uniform asymptotic inference and the bootstrap after model selection.
\newblock {\em The Annals of Statistics\/}~{\em 46\/}(3), 1255--1287.

\bibitem[\protect\citeauthoryear{Tibshirani, Taylor, Lockhart, and
  Tibshirani}{Tibshirani et~al.}{2016}]{Tibshirani.2016}
Tibshirani, R.~J., J.~Taylor, R.~Lockhart, and R.~Tibshirani (2016).
\newblock Exact post-selection inference for sequential regression procedures.
\newblock {\em Journal of the American Statistical Association\/}~{\em
  111\/}(514), 600--620.

\bibitem[\protect\citeauthoryear{Wood}{Wood}{2011}]{mgcv2011}
Wood, S.~N. (2011).
\newblock Fast stable restricted maximum likelihood and marginal likelihood
  estimation of semiparametric generalized linear models.
\newblock {\em Journal of the Royal Statistical Society (B)\/}~{\em 73\/}(1),
  3--36.

\bibitem[\protect\citeauthoryear{Wood}{Wood}{2013}]{Wood.2013}
Wood, S.~N. (2013).
\newblock On p-values for smooth components of an extended generalized additive
  model.
\newblock {\em Biometrika\/}~{\em 100\/}(1), 221--228.

\bibitem[\protect\citeauthoryear{Wood}{Wood}{2017}]{Wood.2017}
Wood, S.~N. (2017).
\newblock {\em Generalized Additive Models: An Introduction with R}.
\newblock CRC press.

\bibitem[\protect\citeauthoryear{Yang, Barber, Jain, and Lafferty}{Yang
  et~al.}{2016}]{Yang.2016}
Yang, F., R.~F. Barber, P.~Jain, and J.~Lafferty (2016).
\newblock Selective inference for group-sparse linear models.
\newblock In {\em Advances in Neural Information Processing Systems}, pp.\
  2469--2477.

\end{thebibliography}

\end{small}

\clearpage

\newgeometry{
  left=1cm,
  right=1cm,
  top=1.5cm,
  bottom=1.5cm
}

\begin{appendix}

\spacingset{1}

\section{Supplementary Material}

\subsection{Further Simulation Results} \label{sec.sim_extra}

\begin{figure}[hbt!]
    \centering
    \includegraphics[width = 0.99\textwidth]{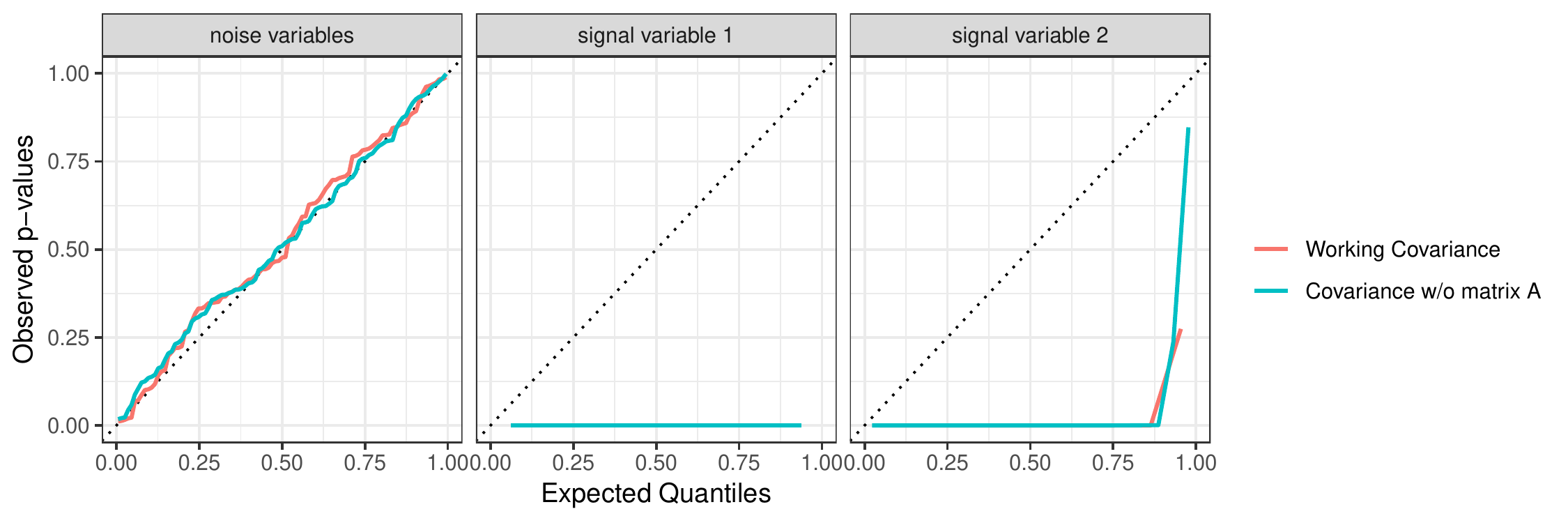}
    \caption{Comparison of expected quantiles of p-values vs. observed (selective) p-values when either using $\bm{G}= \hat{\bm{G}}$ (Working Covariance) or $\bm{G}=\bm{0}_{q \times q}$ (Covariance w/o matrix A) in the definition of the test vector.}
    \label{fig:abc}
\end{figure}

\subsection{Further Details on Section \ref{sec.appl}}\label{sec.AppendixDetails}

\paragraph{Theories of inflation.} The first theory regards the creation of monetary base, credit growth and the Philips Curve. The second theory comprises countries' institutional setup. The third accounts for a group of variables that represent monetary policy strategies while the fourth focuses on public finances. The fifth theoretical explanation is related to globalization, the sixth takes into account demographic changes and the seventh the development of prices and rents of natural resources. The last theory considers the inertia of the inflation process.

\paragraph{Details on model fitting.} For model fitting we use the R-package \texttt{mgcv} \citep{mgcv2011} and obtain model selection via the R-package \texttt{cAIC4} \citep{Saefken.2019}. To allow this, we  extend the \texttt{cAIC4} package to permit calculations for 1) mixed and additive models estimated by \texttt{mgcv} and 2) models beyond i.i.d.\ error covariance following the proposed extension of \citet{overholser2014effective}. As \citet{overholser2014effective} do not take into account the estimation uncertainty of $\bm{G}$, we implemented a working version that adds the number of unknown parameters in $\bm{R}$ (asymptotically justified by  \citet{overholser2014effective}) to the bias correction term of \citet{Saefken.2019}. Further research might strive for an analytic solution to the bias correction term. 

\paragraph{Run-time to derive inference statements for $M^{*}_{4}$.} The run-time for the estimation of the 16 AMMs in $\mathcal{M}_4$ followed by 16 cAIC computations for a single new response vector $\bm{y}^b$ on a cluster of 34 cores (Intel(R) Xeon(R) CPU E3-1284L v4 @ 2.90GHz) was approximately 1:05min. In order to obtain inference statements for a single $\hat{f}_{GDPpc}(z^{+})$ the estimation of all AMMs and cAICs associated with each newly generated response vector $\bm{y}^b$ for  all $B$ samples summed to a total run-time of 26h and 59min.

\paragraph{Test of the bivariate Interaction Effect of $M^{**}$:} In the case of this bivariate interaction effect, ${\bm{W^{*}}}$ is given by $\bm{W^{*}} = \bm{P^{*\bot}_{\bm{X_{A_{9,7}\text{\textbackslash}E_{9,7}}}}} \bm{X_{E_{9,7}}}$ where $\bm{X_{A_{9,7}\text{\textbackslash}E_{9,7}}}$ is the design matrix comprising the evaluated spline basis functions associated with \textit{Credit (\% GDP) Gr.} and with \textit{GDP Gr. (\%)}. $\bm{X_{E_{9,7}}}$ comprises the evaluated basis functions of the bivariate tensor-product spline. We test $H_0: ||\bm{P^{*}}_{\bm{W^{*}}} \bm{\mu}||_2 = 0$ against its one sided alternative $H_A: ||\bm{P^{*}}_{\bm{W^{*}}} \bm{\mu}||_2 > 0$, corresponding to an overall test for the interaction surface. The test statistics is $T^{*} = ||\bm{P^{*}}_{\bm{W^{*}}} \bm{Y_{M^{**}}}||_2$ where $\bm{y_{M^{**}}}$ is the observed realization of $\bm{Y_{M^{**}}}$. As tests for groups of variables are currently only available for unpenalized effects, the degrees of freedom of our selective $\chi$-significance test were chosen according to the rank of $\bm{X_{E_{9,7}}}$, which might lead to a loss in power due to the penalization of the basis coefficients \citep[cf.][]{Wood.2013}. By means of the Monte Carlo approximation described in Section \ref{MC}, we obtained $T^{b^{*}}$ for $b^{*} = 1,\ldots,B^{*} \equiv 4000$. We obtained 80 $T^{b^{*}}$, for which $\mathcal{S}_{sec}(\bm{y_{M^{**}}}) \equiv \mathcal{S}_{sec}(\bm{y^{b^{*}}})$ with $\bm{y^{b^{*}}} = T^{b^{*}} \cdot \text{dir}_{\bm{W^{*}}}(\bm{y_{M^{**}}}) + \bm{P^{*\bot}_{W^{*}}y_{M^{**}}}$. Finally, we obtained a p-value in the same manner as for the selective inference statements for $M_{4}^{*}$ and $M_{5}^{*}$.


\subsection{Tables for Section \ref{sec.appl}}\label{sec.AppendixTables}

\begin{table}[htb!]
\centering
\tiny 
\resizebox{\textwidth}{!}{
\begin{tabular}{l|c|c|c|c|c|c|c|c|c|c|c|c}
\toprule
$z^{+}$  & 2324.2 & 3300.4 & 4538.7 & 4711.0 & 6199.9 & 7852.4 & 8430.3 & 8987.9 & 10999.6 & 13001.1 & 13470.4 & 14023.6 \\
\midrule
\rowcolor{Gray}
  $0 \cdot t_{z^{+}}$ & 11\% & 2\% & 7\% & 14\% & 26\% & 27\% & 27\% & 27\% & 25\% & 6\% & 8\% & 4\% \\  
  $1/3 \cdot t_{z^{+}}$  & 18\% & 6\% & 20\% & 22\% & 26\% & 26\% & 26\% & 26\% & 25\% & 19\% & 15\% & 25\% \\  
  \rowcolor{Gray}
  $2/3 \cdot t_{z^{+}}$& 31\% & 30\% & 32\% & 30\% & 24\% & 25\% & 25\% & 25\% & 25\% & 33\% & 35\% & 35\% \\ 
  $3/3 \cdot t_{z^{+}}$ & 41\% & 62\% & 41\% & 34\% & 23\% & 22\% & 22\% & 22\% & 25\% & 42\% & 41\% & 36\% \\ 
  \midrule
  Total & 759 & 478 & 823 & 1027 & 1345 & 1386 & 1397 & 1401 & 1292 & 868 & 908 & 1045 \\   
  \midrule
  \midrule
  p-value & 0.184 & 0.034 & 0.118 & 0.128 & 0.438 & 0.043 & 0.022 & 0.016 & 0.905 & 0.000 & 0.000 & 0.000 \\ 
\bottomrule
\end{tabular}
}
\caption{Share of $1500$ samples that have led to the original selection for the all twelve $z^{+}$, separated by each component of the Gaussian mixture distribution (leftmost column), with mean values being a ratio of the observed test statistic. The percentages add up to 100\% which is equal to \textit{Total}. The last row contains the p-values that result from testing the null hypothesis.}
\label{Tab:Location4}
\end{table}

\begin{table}[hbt!]
\centering
\tiny
\resizebox{\textwidth}{!}{
\begin{tabular}{l|c|c|c|c|c|c|c|c|c|c|c|c}
\toprule
$z^{++}$   & 25.0 & 28.4 & 31.8 & 60.0 & 80.0 & 120.0 & 169.0 & 183.0 & 200.0 & 260.9 & 311.9 & 362.8 \\ 
\midrule
\rowcolor{Gray}
  $0 \cdot t_{z^{++}}$ & 19\% & 18\% & 21\% & 0\% & 25\% & 0\% & 9\% & 28\% & 42\% & 12\% & 20\% & 22\% \\
  $1/3 \cdot t_{z^{++}}$ & 21\% & 20\% & 21\% & 1\% & 25\% & 10\% & 0\% & 3\% & 11\% & 20\% & 25\% & 26\% \\ 
  \rowcolor{Gray}
  $2/3 \cdot t_{z^{++}}$ & 31\% & 31\% & 29\% & 32\% & 25\% & 40\% & 17\% & 12\% & 9\% & 30\% & 27\% & 27\% \\ 
  $3/3 \cdot t_{z^{++}}$& 29\% & 32\% & 29\% & 67\% & 26\% & 50\% & 74\% & 58\% & 38\% & 38\% & 28\% & 25\% \\  
  \midrule
  Total & 756 & 707 & 755 & 271 & 1059 & 425 & 248 & 347 & 560 & 709 & 971 & 992 \\ 
  \midrule
  \midrule
  p-value & 0.002 & 0.002 & 0.001 & 0.005 & 0.426 & 0.002 & 0.000 & 0.000 & 0.000 & 0.002 & 0.015 & 0.057 \\ 
 \bottomrule
\end{tabular}
}
\caption{Share of 1200 generated samples that have led to the original model selection for each $z^{++}$, separated by each component of the Gaussian mixture distribution which has been the proposal distribution in the importance sampling procedure. The percentages add up to 100\% which is equal to \textit{Total}. The last row contains the p-values that result from testing the null hypothesis.}
\label{Tab:Location5}
\end{table}


\begin{table}[hbt!]
\centering
\resizebox{\linewidth}{!}{%
\tiny 
\begin{tabularx}{\textwidth}{p{0.7cm}P{6cm}P{6cm}P{3cm}P{1.8cm}}
\toprule
 & $\bm{B_{j,l}}$ & $\bm{C_{j,l}}$ & $\bm{D_{j,l}}$ & $\bm{E_{j,l}}$ \\ 
  \midrule
  \rowcolor{Gray}
  $A_{1,1}$ & Output Gap (\%) & M2 Growth (\%), Year & & \\ 
  $A_{2,1}$ & Real GDP Growth (\%) & M2 Growth (\%), Year & & \\ 
  \rowcolor{Gray}
  $A_{3,1}$ & & M2 Growth (\%), Real GDP per capita (USD), Year & &\\ 
  $A_{4,1}$ & Output Gap (\%) & Credit (\% of GDP) Growth, Year &  &  \\ 
  \rowcolor{Gray}
  $A_{5,1}$ & Real GDP Growth (\%) & Credit (\% of GDP) Growth, Year &  & \\ 
  $A_{6,1}$ & & Credit (\% of GDP) Growth, Real GDP per capita (USD), Year &  & \\ 
  \midrule
    \rowcolor{Gray}
  $A_{1,2}$ & M2 Growth (\%), Central Bank Transparency & Real GDP Growth (\%), Year &  &  \\ 
  $A_{2,2}$ & M2 Growth (\%), Real GDP per capita (USD) & Central Bank Transparency, Year &  &  \\ 
        \rowcolor{Gray}
  $A_{3,2}$ & Credit (\% of GDP) Growth, Central Bank Transparency & Real GDP Growth (\%), Year & & \\ 
  $A_{4,2}$ & Credit (\% of GDP) Growth, Central Bank Transparency, Real GDP per capita (USD) & Year & &\\ 
        \rowcolor{Gray}
  $A_{5,2}$ & Political Orientation, Political Stability,  M2 Growth (\%) & Real GDP Growth (\%), Year & Political Orientation / Political Stability & \\ 
  $A_{6,2}$ & Political Orientation, Political Stability,  M2 Growth (\%), Real GDP per capita (USD) & Year & Political Orientation / Political Stability & \\ 
        \rowcolor{Gray}
  $A_{7,2}$ & Political Orientation, Political Stability,  Credit (\% of GDP) Growth, Real GDP Growth (\%) & Year & Political Orientation / Political Stability &  \\ 
  $A_{8,2}$ & Political Orientation, Political Stability,  Credit (\% of GDP) Growth, Real GDP per capita (USD) & Year & Political Orientation / Political Stability &  \\ 
        \rowcolor{Gray}
  $A_{9,2}$  & M2 Growth (\%), Political Rights & Real GDP Growth (\%), Year &  & \\ 
  $A_{10,2}$ & M2 Growth (\%), Political Rights, Real GDP per capita (USD) & Year &  &  \\ 
        \rowcolor{Gray}
  $A_{11,2}$ & Political Rights, Credit (\% of GDP) Growth & Real GDP Growth (\%), Year &  & \\ 
  $A_{12,2}$ & Political Rights, Credit (\% of GDP) Growth, Real GDP per capita (USD), Credit (\% of GDP) Growth & Year &  & \\ 
        \rowcolor{Gray}
  $A_{13,2}$ & Civil Liberty, M2 Growth (\%) & Real GDP Growth (\%), Year &  & \\ 
  $A_{14,2}$ & Civil Liberty, M2 Growth (\%), Real GDP per capita (USD) & Year &  & \\ 
        \rowcolor{Gray}
  $A_{15,2}$ & Civil Liberty, Credit (\% of GDP) Growth & Real GDP Growth (\%), Year &  &  \\ 
  $A_{16,2}$ & Civil Liberty, Credit (\% of GDP) Growth, Real GDP per capita (USD) & Year &  &  \\ 
        \rowcolor{Gray}
  $A_{17,2}$ & Freedom Status, M2 Growth (\%) & Real GDP Growth (\%), Year &  & \\ 
  $A_{18,2}$ & Freedom Status, M2 Growth (\%), Real GDP per capita (USD) & Year &  & \\ 
        \rowcolor{Gray}
  $A_{19,2}$ & Freedom Status, Credit (\% of GDP) Growth, Real GDP per capita (USD) & Year &  & \\ 
  $A_{20,2}$ & Freedom Status, Credit (\% of GDP) Growth & Real GDP Growth (\%), Year &  & \\ 
        \rowcolor{Gray}
  $A_{21,2}$ & M2 Growth (\%), Central Bank Independence, Turn Over Rate & Real GDP Growth (\%), Year & Turn Over Rate / Central Bank Independence & \\ 
  $A_{22,2}$ & M2 Growth (\%), Central Bank Independence, Turn Over Rate, Real GDP per capita (USD) & Year & Turn Over Rate / Central Bank Independence & \\ 
        \rowcolor{Gray}
  $A_{23,2}$ & Credit (\% of GDP) Growth, Central Bank Independence, Turn Over Rate & Real GDP Growth (\%), Year & Turn Over Rate / Central Bank Independence & \\ 
  $A_{24,2}$ & Credit (\% of GDP) Growth, Central Bank Independence, Turn Over Rate, Real GDP per capita (USD) & Year & Turn Over Rate / Central Bank Independence & \\ 
        \midrule
        \rowcolor{Gray}
  $A_{1,3}$ & Exchange Rate Arrangement, Real GDP Growth (\%) & M2 Growth (\%), Year & & \\ 
  $A_{2,3}$ & Exchange Rate Arrangement & M2 Growth (\%), Real GDP per capita (USD), Year & & \\ 
        \rowcolor{Gray}
  $A_{3,3}$ & Exchange Rate Arrangement, Credit (\% of GDP) Growth, Real GDP Growth (\%) & Year & & \\ 
  $A_{4,3}$ & Exchange Rate Arrangement, Credit (\% of GDP) Growth& Real GDP per capita (USD), Year & & \\ 
        \rowcolor{Gray}
  $A_{5,3}$ & Inflation Targeting, Real GDP Growth (\%) & M2 Growth (\%), Year & & \\ 
  $A_{6,3}$ & Inflation Targeting & Real GDP per capita (USD), M2 Growth (\%), Year & & \\ 
        \rowcolor{Gray}
  $A_{7,3}$ & Inflation Targeting, Real GDP Growth (\%), Credit (\% of GDP) Growth & Year & &  \\ 
  $A_{8,3}$ & Inflation Targeting, Credit (\% of GDP) Growth & Real GDP per capita (USD), Year & &  \\ 
        \midrule
  \rowcolor{Gray}
  $A_{1,4}$ & Debt (\% of GDP) Growth & M2 Growth (\%), Real GDP Growth (\%), Year & & \\ 
  $A_{2,4}$ & Debt (\% of GDP) Growth & M2 Growth (\%), Real GDP per capita (USD), Year & & \\ 
      \rowcolor{Gray}
  $A_{3,4}$ & Credit (\% of GDP) Growth, Debt (\% of GDP) Growth & Real GDP Growth (\%), Year & & \\ 
  $A_{4,4}$ & Credit (\% of GDP) Growth,  Debt (\% of GDP) Growth & Real GDP per capita (USD), Year & & \\ 
      \rowcolor{Gray}
  $A_{5,4}$ & Primary Balance (\% of GDP) & M2 Growth (\%), Real GDP Growth (\%), Year & & \\ 
  $A_{6,4}$ & Primary Balance (\% of GDP) & M2 Growth (\%), Real GDP per capita (USD), Year & & \\ 
      \rowcolor{Gray}
  $A_{7,4}$ & Primary Balance (\% of GDP), Credit (\% of GDP) Growth & Real GDP Growth (\%), Year & & \\ 
  $A_{8,4}$ & Primary Balance (\% of GDP), Credit (\% of GDP) Growth & Real GDP per capita (USD), Year & & \\ 
      \rowcolor{Gray}
  $A_{9,4}$ & Maturities & M2 Growth (\%), Real GDP Growth (\%), Year & & \\ 
  $A_{10,4}$ & Maturities & M2 Growth (\%), Real GDP per capita (USD), Year & & \\ 
      \rowcolor{Gray}
  $A_{11,4}$ & Credit (\% of GDP) Growth, Maturities & Real GDP Growth (\%), Year &  & \\ 
  $A_{12,4}$ & Credit (\% of GDP) Growth, Maturities & Real GDP per capita (USD), Year &  & \\ 
      \rowcolor{Gray}
  $A_{13,4}$ & & Denomination (\%), M2 Growth (\%), Real GDP Growth (\%), Year & & \\ 
  $A_{14,4}$ & & Denomination (\%), M2 Growth (\%), Real GDP per capita (USD), Year & & \\ 
      \rowcolor{Gray}
  $A_{15,4}$ & Credit (\% of GDP) Growth & Denomination (\%), Real GDP Growth (\%), Year & & \\ 
  $A_{16,4}$ & Credit (\% of GDP) Growth & Denomination (\%), Real GDP per capita (USD), Year & & \\
 \bottomrule
\end{tabularx}
}
\caption{(1/2) Allocation of the predictor set $A_{j,l}$ of the model $M_{j,l}$ to $B_{j,l}$,$C_{j,l}$,
$D_{j,l}$ and $E_{j,l}$} 
\label{Tab:Allocation1}
\end{table}
 
\begin{table}[hbt!]
\centering
\tiny 
\begin{tabularx}{\textwidth}{p{0.7cm}P{5cm}P{5cm}P{3cm}P{3.8cm}}
\toprule
 & $\bm{B_{j,l}}$ & $\bm{C_{j,l}}$ & $\bm{D_{j,l}}$ & $\bm{E_{j,l}}$ \\ 
              \midrule
  \rowcolor{Gray}
  $A_{1,5}$ & Real GDP Growth (\%) & Financial Openness, M2 Growth (\%), Trade Openness (\% of GDP), Year &  & \\ 
  $A_{2,5}$ & Trade Openness (\% of GDP),  Real GDP Growth (\%) & M2 Growth (\%), Financial Openness, Year & Trade Openness (\% of GDP) / Financial Openness &  \\ 
    \rowcolor{Gray}
  $A_{3,5}$ & Real GDP Growth (\%) & M2 Growth (\%), Year &  & Financial Openness / Trade Openness (\% of GDP)\\ 
  $A_{4,5}$ &  & Financial Openness, M2 Growth (\%), Trade Openness (\% of GDP), Real GDP per capita (USD), Year &  &\\ 
    \rowcolor{Gray}
  $A_{5,5}$ & Trade Openness (\% of GDP) & Financial Openness, M2 Growth (\%), Real GDP per capita (USD), Year & Trade Openness (\% of GDP) / Financial Openness & \\ 
  $A_{6,5}$ & & M2 Growth (\%), Real GDP per capita (USD), Year &  & Financial Openness / Trade Openness (\% of GDP)\\ 
    \rowcolor{Gray}
  $A_{7,5}$ & Real GDP Growth (\%) & Credit (\% of GDP) Growth, Financial Openness, Trade Openness (\% of GDP), Year & & \\ $A_{8,5}$ & Real GDP Growth (\%) & Credit (\% of GDP) Growth, Financial Openness, Trade Openness (\% of GDP), Year & & Financial Openness / Trade Openness (\% of GDP) \\
    \rowcolor{Gray}
$A_{9,5}$ & Real GDP Growth (\%) & Credit (\% of GDP) Growth, Year & & Financial Openness / Trade Openness (\% of GDP) \\
$A_{10,5}$ & & Real GDP per capita (USD), Credit (\% of GDP) Growth, Financial Openness, Trade Openness (\% of GDP), Year & & Financial Openness / Trade Openness (\% of GDP) \\
    \rowcolor{Gray}
$A_{11,5}$ &  Trade Openness (\% of GDP) & Credit (\% of GDP) Growth, Financial Openness, Real GDP per capita (USD), Year & Trade Openness (\% of GDP) / Financial Openness &  \\ 
$A_{12,5}$ & & Credit (\% of GDP) Growth, Real GDP per capita (USD), Year &  & Financial Openness / Trade Openness (\% of GDP)\\
            \midrule
  \rowcolor{Gray}
$A_{1,6}$ & Age 65 (\%), Real GDP Growth (\%) & M2 Growth (\%), Year & &\\ 
$A_{2,6}$ & Age 65 (\%) & M2 Growth (\%), Real GDP per capita (USD), Year & &\\ 
    \rowcolor{Gray}
$A_{3,6}$ & Age 65 (\%), Real GDP Growth (\%) & Credit (\% of GDP) Growth, Year & & \\ 
$A_{4,6}$ & Age 65 (\%), Real GDP per capita (USD) & Credit (\% of GDP) Growth, Year & & \\ 
    \rowcolor{Gray}
$A_{5,6}$ & Age 75 (\%), Real GDP Growth (\%) & M2 Growth (\%), Year & & \\ 
$A_{6,6}$ & Age 75 (\%) & M2 Growth (\%), Real GDP per capita (USD), Year & & \\ 
    \rowcolor{Gray}
$A_{7,6}$ & Age 75 (\%), Real GDP Growth (\%) & Credit (\% of GDP) Growth, Year & & \\ 
$A_{8,6}$ & Age 75 (\%) & Credit (\% of GDP) Growth, Real GDP per capita (USD), Year & &\\ 
          \midrule
  \rowcolor{Gray}
$A_{1,7}$ & Real GDP Growth (\%) & Energy Prices (USD), Energy Rents (\% of GDP), M2 Growth (\%), Year &  & \\ 
$A_{2,7}$ & Real GDP Growth (\%) & Energy Prices (USD), Energy Rents (\% of GDP), M2 Growth (\%), Year &  & Energy Prices (USD) / Energy Rents (\% of GDP) \\ 
        \rowcolor{Gray}
$A_{3,7}$ & Real GDP Growth (\%) & M2 Growth (\%), Year &  & Energy Prices (USD) / Energy Rents (\% of GDP) \\ 
$A_{4,7}$ & & Energy Prices (USD), Energy Rents (\% of GDP), M2 Growth (\%), Real GDP per capita (USD), Year &  &  \\ 
        \rowcolor{Gray}
$A_{5,7}$ & & Energy Prices (USD), Energy Rents (\% of GDP), M2 Growth (\%), Real GDP per capita (USD), Year & &Energy Prices (USD) / Energy Rents (\% of GDP) \\ 
$A_{6,7}$ & & M2 Growth (\%), Real GDP per capita (USD), Year & & Energy Prices (USD) / Energy Rents (\% of GDP) \\ 
        \rowcolor{Gray}
$A_{7,7}$  & Real GDP Growth (\%) & Energy Prices (USD), Energy Rents (\% of GDP), Credit (\% of GDP) Growth, Year &  & \\ 
$A_{8,7}$  & Real GDP Growth (\%) & Energy Prices (USD), Energy Rents (\% of GDP), Credit (\% of GDP) Growth, Year &  & Energy Prices (USD) / Energy Rents (\% of GDP) \\ 
        \rowcolor{Gray}
$A_{9,7}$  & Real GDP Growth (\%) & Credit (\% of GDP) Growth, Year &  & Energy Prices (USD) / Energy Rents (\% of GDP) \\ 
$A_{10,7}$ & & Energy Prices (USD), Energy Rents (\% of GDP), Credit (\% of GDP) Growth, Real GDP per capita (USD), Year &  &  \\ 
        \rowcolor{Gray}
$A_{11,7}$ & & Energy Prices (USD), Energy Rents (\% of GDP), Credit (\% of GDP) Growth, Real GDP per capita (USD), Year & &Energy Prices (USD) / Energy Rents (\% of GDP) \\
$A_{12,7}$ & & Credit (\% of GDP) Growth, Real GDP per capita (USD), Year & & Energy Prices (USD) / Energy Rents (\% of GDP) \\ 
        \midrule
  \rowcolor{Gray}
$A_{1,8}$ & Real GDP Growth (\%) & M2 Growth (\%), Past Inflation (\%), Year & & \\ 
$A_{2,8}$ & & M2 Growth (\%), Past Inflation (\%), Real GDP per capita (USD), Year & & \\ 
      \rowcolor{Gray}
$A_{3,8}$ & Real GDP Growth (\%) & Credit (\% of GDP) Growth, Past Inflation (\%), Year & & \\ 
$A_{4,8}$ & Real GDP per capita (USD) & Credit (\% of GDP) Growth, Past Inflation (\%), Year & & \\ 
\bottomrule
\end{tabularx}
\caption{(2/2) Allocation of the predictor set $A_{j,l}$ of the model $M_{j,l}$ to $B_{j,l}$,$C_{j,l}$,
$D_{j,l}$ and $E_{j,l}$} 
\label{Tab:Allocation2}
\end{table}

\end{appendix}

\end{document}